\documentclass[prd,12pt,reprint, twocolumn,noshowpacs,nofootinbib,amsmath,amssymb,superscriptaddress,preprintnumbers]{revtex4-1}
\usepackage{graphicx,color}
\usepackage{caption}
\usepackage{amssymb,amsmath,mathrsfs}
\usepackage{hyperref}
\newcommand{\be}{\begin{eqnarray}}
\newcommand{\ee}{\end{eqnarray}}
\newcommand{\met}{p_T\!\!\!\!\!\!\slash \;\;}

\begin{document}
\title{Closing in on singly charged scalars}
\author{Snehadri Das}
\affiliation{Department of Physics, Williams College, Williamstown, MA 01267, USA}
\author{Will Howe}
\affiliation{Department of Physics, Williams College, Williamstown, MA 01267, USA}
\author{Brian Shuve}
\affiliation{Harvey Mudd College, 301 Platt Blvd., Claremont, CA 91711, USA}
\author{David Tucker-Smith}
\affiliation{Department of Physics, Williams College, Williamstown, MA 01267, USA}
\author{Ruby Yager}
\affiliation{Department of Physics, Williams College, Williamstown, MA 01267, USA}

\date{\today}

\begin{abstract}
We investigate current experimental constraints and  future search prospects for a hypothetical $SU(2)_w$ and $SU(3)_c$ singlet spin-zero particle that carries unit electric charge: a singly charged scalar (SCS).  In addition to providing useful benchmarks for collider searches, SCS particles are also well-motivated ingredients in relatively minimal dark sectors.   We focus on scenarios in which the SCS decays promptly at colliders to a lepton plus either a neutrino or an invisible  dark-sector particle of negligible mass.  A promptly decaying SCS can easily have appreciable branching ratios to more than one lepton flavor while remaining consistent with constraints on lepton flavor violation.  This broadens the allowed range of SCS masses to extend well beyond those for right-handed selectrons, smuons, or staus.  For particular benchmark SCS branching ratios, we find that SCS masses above $\sim$185 GeV and in  a lower-mass window $\sim80-125$~GeV are still allowed at 95\%~confidence level.  We carry out Monte Carlo simulations to explore the potential of a boosted-decision-tree-based analysis to probe the surviving SCS parameter space in future searches at the (HL-)LHC, finding a significant increase in sensitivity relative to cut-based analyses both in the lower-mass window and at higher SCS masses.    
\end{abstract}
\maketitle

\section{Introduction}\label{sec:intro}

The Large Hadron Collider (LHC) has proven to be a remarkably powerful probe of physics beyond the Standard Model (SM).  So far, this has meant strong constraints on beyond-the-SM (BSM) scenarios, particularly those with missing-energy signatures.  Using supersymmetry (SUSY) as a benchmark,  masses beyond 2 TeV have been probed for gluinos \cite{ATLAS:2020syg,CMS:2019zmd} and up to about 1 TeV for electroweak charged superpartners \cite{ATLAS:2021yqv, CMS:2022sfi}.   In this paper, we focus on the phenomenology of a hypothetical singly charged scalar (SCS), a BSM particle whose properties are much less constrained \cite{Cao:2017ffm, Babu:2019mfe, Crivellin:2020klg}. We aim to motivate  collider searches dedicated to probing the SCS parameter space to the fullest extent possible, including currently untested regions that extend below 100 GeV in SCS mass.  

An SCS is an $SU(2)_w$ and $SU(3)_c$ singlet spin-zero particle that carries unit electric charge;  that is, it has the same SM quantum numbers as a right-handed slepton in SUSY.  At the LHC, SCS pair production proceeds via $s$-channel $Z/\gamma$ exchange.  We focus on scenarios in which each SCS decays promptly to a charged lepton plus an invisible particle, leading to events with two opposite-sign leptons and missing transverse momentum.  SCS decay to $l\nu$ can proceed via the renormalizable interaction
\begin{equation}
\label{eq:nu_coupling}
    {\mathcal L} \supset -\frac{\lambda_{\alpha \beta}}{2} l_\alpha l_\beta \Phi^{*}+{\rm h.c.},
\end{equation}
expressed in two-component notation.  Here, $\Phi$ is the SCS field (which we define to carry negative electric charge), $l_\alpha$ is the SM lepton doublet with flavor $\alpha$, and the implied contraction of lepton-doublet $SU(2)_w$ indices requires the $\lambda$ couplings to be anti-symmetric in flavor space.  The SCS  might alternatively decay to $l\chi$, where $\chi$ is a dark-sector SM-singlet fermion, through the renormalizable interaction
\begin{equation}
\label{eq:DM_coupling}
    {\mathcal L} \supset -F_\alpha e^c_\alpha \chi \Phi+{\rm h.c.}
\end{equation}
For sufficiently light $\chi$, the results of our studies apply equally well to this scenario.

In the framework we adopt, SCS collider phenomenology depends on the SCS mass and its branching ratios to the individual lepton flavors, ${\mathcal B}_{e,\mu,\tau}$, which we assume sum to one.  LHC slepton searches~\cite{ATLAS:2024fub,CMS:2022syk,Aad:2019lff,CMS:2020bfa,CMS:2018eqb,ATLAS:2014zve} place constraints on an SCS with ${\mathcal B}_e  = 1$, ${\mathcal B}_\mu  = 1$, or ${\mathcal B}_\tau  = 1$. In these cases, the SCS signals are identical to right-handed selectrons, smuons, or staus, respectively, which decay directly to an effectively massless neutralino (and with all other superpartners decoupled).   Assuming that the $\lambda$ and $F$ couplings are too small to appreciably affect the $e^+ e^-\rightarrow \Phi\Phi^*$ cross section,  LEP constraints on the SCS parameter space can similarly be inferred from the $\tilde{l}_R$ limits given in Ref.~\cite{lepsusywg}.  The 95\%~CL excluded mass ranges for the three scenarios are as follows:
\begin{itemize}
\item
For ${\mathcal B}_e  = 100\%$, LHC searches exclude the mass range $ 107$~GeV$\lesssim M_\Phi\lesssim 425$~GeV~\cite{ATLAS:2014zve,Aad:2019lff}, leaving a small gap from $\sim 100-107$~GeV between LEP and LHC exclusions.  Our estimate of the lower boundary of this mass window is only approximate, because the $e^+e^-\rightarrow {\tilde e}_R{\tilde e}_R^*$ amplitude includes a contribution from $t$-channel neutralino exchange (calculated in Ref.~\cite{lepsusywg} for a particular SUSY model), which is absent in the SCS scenario with sufficiently small $F$ couplings.  

\item 
For ${\mathcal B}_\mu  = 100\%$, the LEP exclusion is $M_\Phi < 97.3$~GeV, which overlaps with the combined LHC exclusion, $93.5 < M_\Phi < 450$~GeV \cite{ATLAS:2014zve, Aad:2019lff}, leaving no gap at 95\%~CL.   

\item 
For ${\mathcal B}_\tau  = 100\%$, the LEP exclusion is $M_\Phi < 90.1$~GeV and the LHC exclusion is $100 \lesssim M_\Phi \lesssim 350$~GeV \cite{ATLAS:2024fub}, leaving a gap from $\sim 90-100$~GeV between LEP and LHC exclusions.
\end{itemize}
These ``single-flavor'' SCS scenarios are useful benchmarks. However, regardless of whether the SCS decays through the interaction of Eq.~(\ref{eq:nu_coupling}) or Eq.~(\ref{eq:DM_coupling}), there is no particular reason why the SCS should only couple to one individual flavor.  In fact, if the $\lambda$ couplings dominate the decay width, the largest any of the three branching ratios can be is 50\%.   As we will see, even for $M_\Phi\sim 100$~GeV, there is a broad range of $\lambda$ coupling strengths, covering at least $\sim 4$ orders of magnitude, that evade constraints from low-energy probes while inducing prompt SCS decays at colliders.  The range is even larger for particular $\lambda$ flavor structures, and it also widens with increasing $M_\Phi$. This makes a promptly decaying SCS with significant branching ratios to multiple lepton flavors an interesting target for future collider searches.  

The outline of the rest of the paper is as follows.  In Section \ref{sec:bench}, we introduce the SCS benchmark branching ratios that we adopt for the rest of the paper. In Section \ref{sec:DM}, we highlight a possible SCS-dark matter (DM) connection, showing in particular that it is consistent for an SCS that decays promptly at colliders to play a direct role in DM freeze-in.   We carefully reinterpret an ATLAS slepton search \cite{Aad:2019lff} for our SCS benchmark in Section \ref{sec:recast}, and we consider constraints from other LHC searches, LEP, and searches for lepton flavor violation in Section \ref{sec:otherconstraints}.  We find that the SCS mass region above $\sim$185 GeV is still allowed at 95\%~CL, along with lower-mass window from $\sim 80-125$~GeV.
Finally, in Section~\ref{sec:BDT}, we estimate the potential of a BDT-based analysis to probe the surviving SCS parameter space in future (HL-)LHC searches, finding, for example, that significant progress addressing the lower-mass window should be possible provided that relatively low lepton-$p_T$ thresholds can be maintained.  

\section{An SCS benchmark}
\label{sec:bench}

For the rest of the paper, we adopt a benchmark SCS scenario with a particular set of branching ratios: ${\mathcal B}_e ={\mathcal B}_\mu = 25\%$ and ${\mathcal B}_\tau = 50\%$. For example, if SCS decays are induced by the $\lambda$ couplings of Eq.~(\ref{eq:nu_coupling}), these  branching ratios are realized by having $\lambda_{e\tau} = \lambda_{\mu\tau}$, both much larger than $\lambda_{e\mu}$.   In Sec.~\ref{sec:recast} we reinterpret an ATLAS search \cite{Aad:2019lff} that does not report the breakdown of same-flavor, opposite-sign dilepton events into $e^+ e^-$ and $\mu^+ \mu^-$ subcategories.  For our benchmark, which has ${\mathcal B}_e = {\mathcal B}_\mu$, it is unlikely that using that additional information would significantly strengthen the constraints we obtain using publicly available information.  Because our benchmark furthermore takes ${\mathcal B}_\tau$ to be larger than both ${\mathcal B}_e$ and ${\mathcal B}_\mu$ (but still significantly less than 100\%),  it is a particularly challenging one to probe experimentally. We adopt it to highlight current gaps in sensitivity, with the understanding that future experimental searches should target the full SCS parameter space.  

\section{Possible connection to freeze-in dark matter}
\label{sec:DM}
The relative simplicity of SCS extensions of the SM makes them useful theory benchmarks for collider studies.  The possible connection to DM provides an additional motivation.  For feeble $F$ couplings, the interaction of Eq.~(\ref{eq:DM_coupling}) can realize freeze-in DM
\cite{McDonald:2001vt,Hall:2009bx,Bernal:2017kxu}, as recently explored in Refs.~\cite{Berman:2022oht,Asadi:2023csb}.  Imposing a $Z_2$ symmetry under which only the BSM particles are odd stabilizes the DM and forbids the interaction of Eq.~\ref{eq:nu_coupling}; in the $Z_2$-preserving case with a single $\chi$ species,  $\Phi$ is sufficiently long-lived for much of the freeze-in parameter space that the relevant collider probes are searches for long-lived particles, rather than the prompt searches that we focus on in this paper. However, the interactions of Eq.~(\ref{eq:DM_coupling}) can realize DM freeze-in via early-universe $\Phi$ decays, even if the $\lambda$ couplings of Eq.~(\ref{eq:nu_coupling}) are large enough to induce prompt $\Phi$ decays at colliders.  In freeze-in models that incorporate baryogenesis,  these $\lambda$ couplings can in fact dramatically enhance the baryon asymmetry, broadening the viable parameter space~\cite{Berman:2022oht}.  

In the freeze-in regime, we match the observed DM energy density when the DM couplings satisfy the constraint
\begin{equation}
    \sqrt{\sum_\alpha |F_
    \alpha|^2} \simeq 1.8\times 10^{-9}\;
    \left( \frac{100\;\text{keV}}{M_\text{DM}}\right)^{1/2}
    \left( \frac{M_\Phi}{100\;\text{GeV}}\right)^{1/2}.
\end{equation}
Turning on the $\lambda$ couplings destabilizes the DM, leading to a potentially observable gamma-ray line signature from $\chi \rightarrow \gamma\nu ,\, \gamma{\overline \nu}$ decays, which have a combined partial width of \cite{Berman:2022oht}
\be\label{eq:DMdecay}
\Gamma
= \frac{\alpha}{512 \pi^4} \frac{M^3_\text{DM}}{M_\Phi^4}
\sum_\alpha
\left|
\sum_\beta
\lambda_{\alpha \beta} F_{\beta} M_\beta
\left(
\log
\frac{M_\Phi^2}{M_\beta^2}
-1
\right)
\right|^2 \!\!,\;\;
\ee
where the $M_\beta$ are the SM lepton masses. For purposes of illustration, we analyze the resulting gamma-ray constraints for a particular arrangement of $Z_2$-violating couplings that realizes the benchmark SCS branching ratios introduced in Section~\ref{sec:bench}: $\lambda_{e\tau} = \lambda_{\mu \tau} = \lambda$, $\lambda_{e\mu} = 0$.  Due to the appearance of the SM lepton masses  in Eq.~(\ref{eq:DMdecay}), the strength of the gamma-ray line signal  depends strongly on the relative sizes of $F_e$, $F_\mu$, and $F_\tau$.  Given a particular flavor structure for these DM couplings, along with values for $M_\Phi$ and $M_\text{DM}$, gamma-ray observations impose an upper bound on $\lambda$ and thus a lower bound on the SCS lifetime.  These constraints are typically stronger than those from lepton flavor violation, which we will return to in Section \ref{sec:otherconstraints}.

\begin{figure*}[t] 
   \centering
\includegraphics[width=0.65\textwidth]{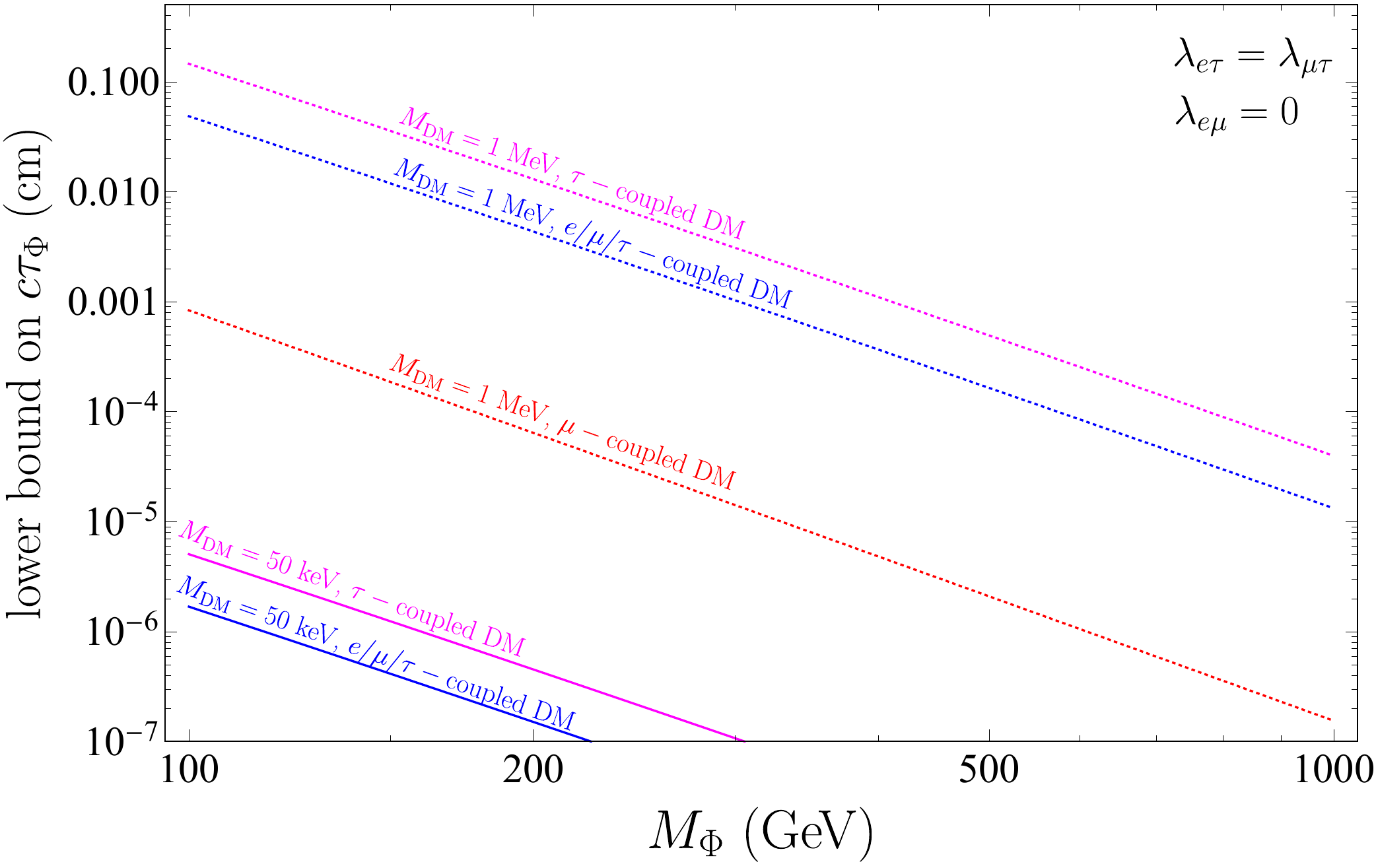}  
   \caption{
For the freeze-in DM scenario described in the text, lower bounds on the SCS lifetime coming from constraints on DM decay, for $M_\text{DM}=50$~keV (solid) and $M_\text{DM}=1$~MeV (dotted). For the $Z_2$-violating couplings of the SCS, we take $\lambda_{e\tau} = \lambda_{\mu\tau}$ and $\lambda_{e\mu}=0$, consistent with the benchmark SCS branching ratios that we adopt for our collider studies: ${\mathcal B}_e ={\mathcal B}_\mu = 25\%$ and ${\mathcal B}_\tau = 50\%$.  We adjust the DM coupling strength to match the observed DM energy density, and we show results for scenarios in which DM couples to $\tau$ (magenta), to all three lepton flavors with equal strength (blue), and to $\mu$ (red).  The constraints for the case of $e$-coupled DM are too weak to appear on the plot, as is the constraint from $\mu \rightarrow e \gamma$. 
  }
   \label{fig:DM_decay_constraints}
\end{figure*}
Fig.~\ref{fig:DM_decay_constraints} shows SCS lifetime lower bounds for $M_\text{DM} = 50$~keV, where the strongest constraint comes from NuSTAR \cite{Perez:2016tcq,Neronov:2016wdd,Ng:2019gch}, and for $M_\text{DM} = 1$~MeV, where the strongest constraint comes from INTEGRAL \cite{Calore:2022pks}.  For $M_\text{DM} = 50$~keV,  the bounds on the SCS lifetime are relatively weak even for the case of $\tau$-coupled DM:  for the full SCS mass range, $c\tau_\Phi$ is allowed to be several orders of magnitude below the $\sim 10^{-2}$~cm upper limit for prompt searches. If the DM couples predominantly to $\mu$ or $e$, the bounds weaken further.  The constraints tighten for heavier DM masses, but for $M_\text{DM} = 1$~MeV, prompt SCS decays are still allowed for the entire SCS mass range for $\mu$- or $e$-coupled DM.  For $M_\text{DM}=1$~GeV, the FERMI and EGRET  lower bounds on $c\tau_\Phi$ inferred from Refs.~\cite{Pullen:2006sy,Essig:2013goa,Albert:2014hwa} are over four orders of magnitude larger than those shown for $M_\text{DM}=1$~MeV, leaving $e$-coupled DM as the only scenario consistent with prompt SCS decays for the entire SCS mass range. 

Eq.~(\ref{eq:DM_coupling}) does not include the possible effects of a neutrino-portal coupling $l\chi H$.  In the absence of this coupling at a high scale ({\em e.g.} the Planck scale), its radiative generation leads to a subdominant correction to the partial width of Eq.~(\ref{eq:DMdecay}) for SCS masses up to $\sim 700$~GeV for the case of $\tau$-coupled DM, and up to higher masses for $e$ or $\mu$ coupled DM \cite{Berman:2022oht}.  The UV-sensitive neutrino-portal-generated contributions to the $\chi \rightarrow \gamma \nu,\,\gamma {\overline \nu}$ amplitudes could conceivably (partially) cancel the IR-dominated contributions we have focused on, weakening the bounds of Fig.~\ref{fig:DM_decay_constraints}. Of course, the SCS might not play a role in DM production at all, in which case  gamma-ray observations are not relevant.  

If $\Phi$ interacts with multiple dark-sector species, prompt $\Phi$ decays can be consistent with freeze-in DM production even in the absence of the $Z_2$-violating $\lambda$ couplings.  For example, $\Phi$ might couple feebly to a heavier state $\chi_2$ that constitutes the observed DM, while coupling somewhat less weakly to a lighter state $\chi_1$ such that $\Phi \rightarrow \chi_1 l$ decays are prompt.  If  $\chi_1$ is light enough ($M_{\chi_1} \lesssim$~eV), it does not contribute appreciably to the present-day DM energy density and evades constraints on light relics, even if it comes into equilibrium in the early universe \cite{Xu:2021rwg}, while the  $\chi_2 \rightarrow \chi_1 \gamma$ decay rate can be sufficiently long for appropriate $\chi_2$ masses and couplings.  

Finally, there is also the possibility of freeze-{\em out} production, the DM scenario for which which the interaction of Eq.~\ref{eq:DM_coupling} was first considered \cite{Agrawal:2011ze,Batell:2013zwa,Bai:2014osa,Chang:2014tea,Agrawal:2014una,Agrawal:2014aoa,Agrawal:2015tfa,Agrawal:2015kje,Agrawal:2016uwf,Desai:2020rwz,Acaroglu:2022hrm}.  For the larger $F$ couplings needed for freeze-out, constraints on lepton flavor violation typically force $\Phi$ to couple to an individual lepton flavor, or else require multiple DM states with DM couplings in approximate flavor alignment with the SM lepton Yukawa couplings.  The latter case can lead to $\Phi$ having appreciable branching ratios to multiple lepton flavors, which is the phenomenological scenario we focus on here.  For simplicity, we restrict our attention in this paper to the case in which the invisible product(s) of $\Phi$ decay have negligible mass as far as collider physics is concerned, whereas larger DM masses would be required for freeze-out.

\section{Reinterpretation of an ATLAS slepton search}
\label{sec:recast}
An SCS with mass $\gtrsim 100$~GeV and appreciable branching ratios to $e$ and/or $\mu$ is constrained by a 13 TeV ATLAS search for direct slepton production in final states with two leptons plus missing transverse momentum, based on an integrated luminosity of 139~fb$^{-1}$~\cite{Aad:2019lff}. The ATLAS collaboration has provided a wealth of auxiliary materials \cite{hd89413} to accompany this analysis via the {\tt HistFactory} framework~\cite{Cranmer:2012sba}, including the observed counts and the expected background counts, along with (correlated) systematic uncertainties, in 36 signal regions used for their binned analysis.   These signal regions incorporate same-flavor, opposite-sign dilepton events (SF: $e^+ e^-$ or $\mu^+ \mu^-$) and  different-flavor opposite-sign dilepton events  (DF: $e^+ \mu^-$ or $\mu^+ e^-$), with either zero jets (0J) or one jet (1J).  Each of the four categories (SF0J, SF1J, DF0J, DF1J) is divided into nine $M_{T_2}$ \cite{Lester:1999tx,Barr:2003rg} bins\footnote{We define $M_{T_2}$ to be calculated with the mass of the invisible particle set to zero.}, giving 36 signal regions in all.    Using the information provided, it is possible to estimate constraints on any BSM model based on the expected counts in these 36 signal-regions found in MC simulations.

To reinterpret the results of Ref.~\cite{Aad:2019lff} for an SCS, we perform Monte Carlo simulations of SCS pair production at the LHC to estimate efficiencies for the 36 ATLAS signal regions, for given sets of SCS parameters.  We use FeynRules~\cite{Alloul:2013bka} to create the appropriate model files, and carry out these simulations using a  MadGraph/MadEvent \cite{Alwall:2014hca}$\rightarrow$Pythia\cite{Bierlich:2022pfr}$\rightarrow$Delphes\cite{deFavereau:2013fsa} event-generation pipeline. Working within the MadAnalysis framework \cite{Conte:2012fm},  the authors of Ref.~\cite{Araz:2020dlf} developed an ``offline'' analysis code that approximates the ATLAS analysis (Ref.~\cite{Aad:2019lff}) when applied in tandem with a customized Delphes card.  This  code includes treatments of lepton isolation and unique object identification designed to be consistent with the actual experimental search.   Ref.~\cite{Araz:2020dlf}  shows that this MadAnalysis implementation reproduces to a good approximation the simulated signal efficiencies reported by ATLAS for direct slepton pair production within a particular benchmark simplified SUSY model.  We adopt the same customized Delphes card and analysis code with a few modifications described in Appendix~\ref{sec:MC}, where we also provide additional details about our Monte Carlo simulations and the cutflow from Ref.~\cite{Aad:2019lff}.    
\begin{figure}[t] 
   \centering
   \includegraphics[width=3.5in]{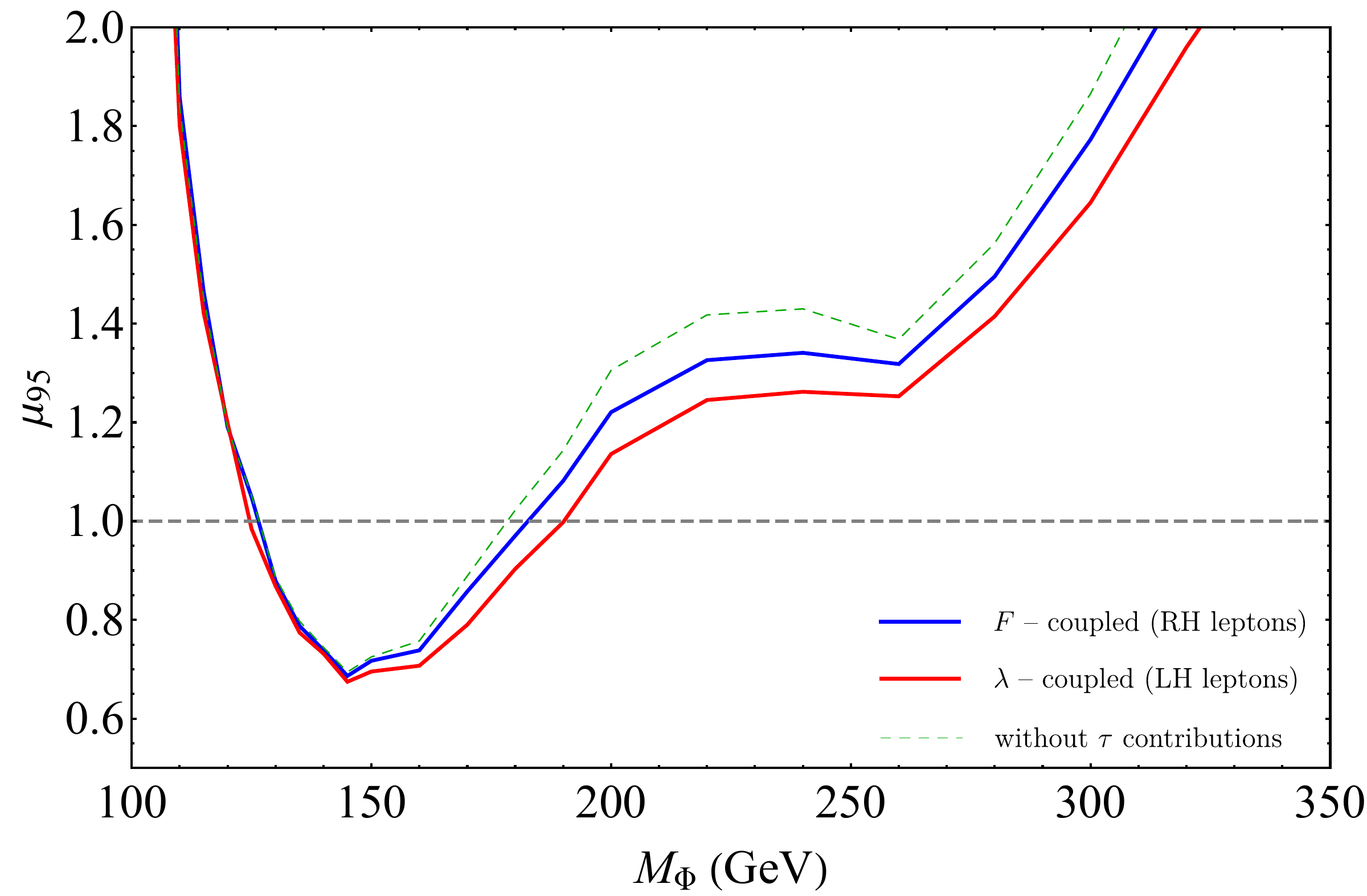} 
   \caption{Reinterpretation of Ref.~\cite{Aad:2019lff}, the ATLAS search for direct slepton production, for an SCS with ${\mathcal B}_e = {\mathcal B}_\mu = 25\%$ and ${\mathcal B}_\tau = 50\%$.   The contours give $\mu_{95}$, the 95\% CL upper limit on the signal strength, as a function of SCS mass;  masses with $\mu_{95} < 1$ are ruled out at 95\% CL for the chosen branching ratios.  The blue and red contours apply, respectively, for an SCS that decays through the DM coupling of Eq.~(\ref{eq:DM_coupling}), and for an SCS decays through the neutrino coupling of Eq.~(\ref{eq:nu_coupling}).  The two  scenarios give slightly different results because the helicity of $\tau$s produced in SCS decays impacts the $e/\mu$ spectrum from leptonic $\tau$ decays.  Not including events with SCS decays to $\tau$  gives the dashed, green contour. 
   }
   \label{fig:recast_plot}
\end{figure}
The LHC cross sections for SCS pair production are the same as for direct production of right-handed sleptons, which we take from Refs.~\cite{susyxsecwg,Bozzi:2007qr,Fuks:2013vua,Fuks:2013lya,Fiaschi:2018xdm,Beenakker:1999xh}. 
With all 36 signal-region expected signal counts in hand for a given SCS model, we use the {\tt HistFactory} information provided by ATLAS and {\tt pyhf} \cite{Heinrich:2021gyp} to calculate CL$_\text{s}$ values~\cite{Read:2002hq} and 95\%~CL signal-strength upper limits for that model.  In Appendix~\ref{sec:recast_check}, we apply this procedure to our own simulated slepton signals in order  to directly compare with ATLAS results, and we find reasonable agreement.  

We present signal-strength upper limits for an SCS with our benchmark branching ratios in Figure~\ref{fig:recast_plot}, which shows that only the mass range $125\text{ GeV}\lesssim M_\Phi \lesssim 185\text{ GeV} $ is excluded at 95\% CL.  The relatively weak limits on an SCS with appreciable ratios to more than one lepton flavor motivate future dedicated collider searches for this scenario.  


We obtain stronger limits than the LHC constraints from Ref.~\cite{Cao:2017ffm}, which used  Run I results, and from Ref.~\cite{Babu:2019mfe}, which used a Run II CMS search  based on an integrated luminosity of  36~fb$^{-1}$ \cite{CMS:2018eqb}.   Our results are roughly consistent with the lower-bound estimate  $M_\Phi \gtrsim 200$~GeV obtained in Ref.~\cite{Crivellin:2020klg} using the same ATLAS search we consider here.  The authors of Ref.~\cite{Crivellin:2020klg} arrived at their estimate by using cross-section times branching ratio alone ({\em i.e.} without simulating signal efficiencies) and considering only same-flavor events. Their estimate does not fully capture the $M_\Phi$ dependence of the signal efficiency, which opens a window at lower masses, $M_\Phi \lesssim 125$ GeV. 

We present results for other simple sets of branching ratios in Appendix~\ref{sec:recast_various_branchings}.  For a two-flavor scenario with  ${\mathcal B}_\mu ={\mathcal B}_\tau = 50\%$, for example, we estimate that the mass range excluded by the ATLAS search expands to $120\text{ GeV}\lesssim M_\Phi \lesssim 255\text{ GeV}$.

In our analysis, we do take into account the contribution to the signal from SCS decays to $\tau$ leptons, which can subsequently decay leptonically. The leptons produced by $\tau$ decays are softer than those produced by direct SCS decays, which is part of the reason this contribution has a relatively modest impact.   The light-flavor leptons produced in $\tau$ decays tend to be somewhat harder in the scenario in which the SCS decays through the $\lambda$ couplings of Eq.~(\ref{eq:nu_coupling}), because in that case the decaying $\tau$ and the light lepton typically have the same helicity.  That leads to slightly stronger limits in the $\lambda$-coupled scenario than in the $F$-coupled scenario, as shown in Fig.~\ref{fig:recast_plot}.

\section{Other SCS constraints}
\label{sec:otherconstraints}

In this section, we summarize other experimental constraints on SCS particles. Ultimately, we conclude that the disallowed masses for our SCS benchmark are the $\simeq 125-185$~GeV excluded range already identified in our reinterpretation of Ref.~\cite{Aad:2019lff}, and $M_\Phi \lesssim 80$~GeV, based on LEP results.

\subsection{Other LHC searches for selectrons and smuons}

A CMS search \cite{CMS:2020bfa} based on 137~fb$^{-1}$ places constraints on direct slepton pair production that are comparable to  those obtained by the ATLAS search used in Sec.~\ref{sec:recast}, Ref.~\cite{Aad:2019lff}.  For example, both searches place an upper limit of $\sim 2$~fb on the total production cross section for 200-GeV degenerate sleptons (${\tilde e}_{L,R}$ and ${\tilde \mu}_{L,R}$) decaying to leptons plus   a massless neutralino.  The CMS analysis defines signal regions to allow SF events only, while DF events are steered into control regions.
This would appear to reduce the sensitivity of the CMS search to our SCS benchmark.  More recently, a CMS Physics Analysis Summary~\cite{CMS-PAS-SUS-23-002} presents preliminary results on a search for new physics in events with oppositely charged leptons and missing transverse momentum, using both SF and DF control regions.  We have not attempted to carry out a detailed reinterpretation of this study, but we have checked that if we estimate the SCS to have a signal efficiency smaller than that for degenerate sleptons by a factor of four (to approximately take into account, for our SCS benchmark,  the requirement of two light leptons, whether SF or DF), the signal-strength upper limits come out to 2.7 for $M_\Phi = 120$~GeV and 2.3 for $M_\Phi = 180$~GeV, suggesting that this analysis is not likely to shrink the allowed mass ranges identified in Fig.~\ref{fig:recast_plot}.  For these reasons, and because the supplementary materials provided with  Ref.~\cite{Aad:2019lff} make that analysis particularly straightforward to reinterpret for an SCS that can decay to multiple flavors,  we have chosen to focus on the ATLAS search.     

In the lower-mass region, an ATLAS 8~TeV search \cite{ATLAS:2014zve} places stronger constraints on right-handed sleptons than the 13~TeV search ATLAS search, excluding ${\tilde e}_R$ and ${\tilde \mu}_R$ for masses down to 107.4~GeV and 93.5~GeV, respectively, for the case of a massless neutralino \cite{hd63216}.  To obtain a rough estimate of signal-strength upper limits for our SCS benchmark, we again approximate the cross section upper limit for the SCS to be the reported  cross section upper limit for combined ${\tilde e}_R$ and ${\tilde \mu}_R$ production divided by four, because only about a quarter of the signal events have two light leptons (and DF events are included in the signal regions for this analysis).  Focusing on the lower-mass region, this leads to signal-strength upper limits of $\simeq 2.4$ and 4.5 for SCS masses of 125~GeV and 100 GeV, respectively.  Comparing with Fig.~\ref{fig:recast_plot}, the constraint we obtained from the 13~TeV search is stronger at 125~GeV.  The 8~TeV constraint is stronger at 100~GeV, but does not qualitatively change the basic point that there exists surviving parameter space to be probed.  Similarly,  an ATLAS Run-II slepton search based on 36.1~fb$^{-1}$, Ref.~\cite{ATLAS:2018ojr}, does not further impact the parameter space for our SCS benchmark.  Using the auxiliary material from 
Ref.~\cite{81996hd} 
to perform the identical exercise as for the ATLAS 8~TeV search, we estimate signal strength upper limits of $\sim 2$ and $\sim30$ for SCS masses of 150~GeV and 100 GeV, respectively.

Finally, the 13~TeV ATLAS search described in Ref.~\cite{ATLAS:2022hbt} targets charginos and sleptons decaying to leptons plus neutralino with a mass splitting comparable to the $W$ mass. The slepton search includes only SF events in the signal region.  Using the auxiliary material available at the associated HEPData repository \cite{hd134068},  we conclude that this search is not competitive with the search of Ref.~\cite{Aad:2019lff} at higher SCS masses.  For example, the reported cross-section upper limit for direct production of degenerate sleptons (${\tilde e}_{L,R}$ and ${\tilde \mu}_{L,R}$) decaying to lepton plus massless neutrino is 11.9~fb for a slepton mass of 200 GeV, about six times larger than the corresponding limit for Ref.~\cite{Aad:2019lff}.  Focusing on our SCS benchmark at lower masses, we can obtain a very rough estimate for the signal-strength upper limit by dividing the reported cross-section upper limit for degenerate sleptons  by one-eighth of the SCS cross section, with the rationale than only that fraction of signal events has same-flavor light leptons for our benchmark.  This leads to signal strength upper limits of $\simeq 5.9$ and $\simeq 4.7$ at SCS masses of 125 GeV and 100~GeV, respectively.  Although these estimates are highly approximate, as they neglect efficiency differences for left and right-handed slepton production, for example, they strongly suggest that Ref.~\cite{ATLAS:2022hbt} does not improve upon the constraints on our SCS benchmark shown Fig.~\ref{fig:recast_plot}.  

\subsection{LHC searches for staus}
CMS~\cite{CMS:2022syk} and ATLAS~\cite{ATLAS:2024fub} searches for direct stau production using the full Run II datasets obtain 95\% CL exclusions for  pure $\tilde{\tau}_R$ pair-production, with $\tilde{\tau}_R$ decaying promptly to tau plus an essentially massless neutralino,  in the $\sim 150-250$ GeV and $\sim 100-350$ GeV mass ranges, respectively.  For our benchmark SCS scenario, which has ${\mathcal B}_\tau = 50\%$, neither search excludes any SCS masses.  Taking the signal cross section to come with an extra suppression of $1/4$ for this benchmark, we estimate that the somewhat more constraining ATLAS search gives 95\% CL signal-strength upper limits of $\simeq$~3.6, 2.1, 1.5, and 2.0 for $M_\Phi = $~100, 120, 160, and 200~GeV, respectively.  Aside for the $M_\Phi = 100$~GeV upper limit, these are less stringent than those based on the ATLAS dilepton search, shown in Fig~\ref{fig:recast_plot}.  

\subsection{LEP constraints}
For pair production of right-handed sleptons decaying to lepton plus a massless neutralino, 
LEP searches \cite{lepsusywg} establish the lower mass bounds of 100~GeV for $\tilde{e}_R$, 97~GeV for $\tilde{\mu}_R$, and 90~GeV for $\tilde{\tau}_R$, which can be interpreted as SCS mass lower bounds for $\mathcal{B}_{e,\mu,\tau}=100\%$, respectively.  The limits on stau production turn out to be the most constraining for our SCS benchmark, which has ${\mathcal B}_\tau$ twice as large as ${\mathcal B}_e$ and ${\mathcal B}_\mu$.  Calculating the leading-order cross section for $e^+ e^- \rightarrow \tilde{l}_R^+ \tilde{l}_R^-$ at $\sqrt{s} = 208$~GeV with MadGraph/MadEvent~\cite{Alwall:2014hca}, and using the available 95\%~CL cross-section upper limits \cite{lepsusywg}, we estimate that the limits on ${\tilde \tau}_R$ production constrain the SCS mass to be above $\simeq 80$~GeV for our SCS benchmark. Based on the information available, the  ${\tilde \mu}_R$ constraint allows masses going down to at least $\simeq 60$~GeV (and perhaps lower), while masses all the way down to $45$~GeV appear to be allowed by the ${\tilde e}_R$ constraint.  Taking into account both LEP and LHC constraints, the allowed lower-mass range for our SCS benchmark is therefore $\simeq 80-125$~GeV. 

Our 80-GeV estimate of the LEP reach is slightly higher than the $\sim 70$~GeV value obtained in Ref.~\cite{Cao:2017ffm} for $\mathcal{B}_e+\mathcal{B}_\mu = 0.5$.  The authors of Ref.~\cite{Cao:2017ffm} allow the SCS to have an appreciable branching ratio to quarks via higher-dimension operators, a possibility we choose not to explore for simplicity.  The associated operators can be forbidden by imposing a generalized lepton number symmetry under which the SCS is charged.  

\subsection{Low-energy probes and precison electroweak data}
\label{sec:LFV}
For an SCS with a mass in the $\gtrsim 100$~GeV range there exists a broad range of coupling strengths that are small enough to evade constraints on lepton flavor violation  and other low-energy processes, and large enough that the SCS qualifies as promptly decaying for collider physics purposes.  As discussed in Ref.~\cite{Berman:2022oht},  bounds from modifications of the muon decay rate, $\mu\rightarrow e \gamma$, and other low-energy phenomena can be determined for the SCS scenario with the couplings
\begin{equation}
    {\mathcal L} \supset -\frac{\lambda_{\alpha \beta}}{2} l_\alpha l_\beta \Phi^{*}+{\rm h.c.}
\end{equation}
by appropriately adapting results obtained in the context of R-parity-violating supersymmetry \cite{Barbier:2004ez}.  As a concrete example, we return to the scenario with $\lambda_{e\tau} = \lambda_{\mu \tau} = \lambda$ and $\lambda_{e \mu} = 0$, which realizes the SCS benchmark branching ratios of Sec.~\ref{sec:bench}.  The strongest constraint then comes from $\mu\rightarrow e \gamma$; the experimental upper limit on the branching ratio obtained by Ref.~\cite{MEG:2016leq} implies
\begin{equation}
\label{eq:mutoegamma}
    |\lambda| \lesssim 6 \times 10^{-3} \left(\frac{M_\Phi}{100\rm{\;GeV}} \right).
\end{equation}
Meanwhile, the $\Phi$ lifetime can be expressed as
\begin{equation}
    c\tau_\Phi =
\left(\frac{100\rm{\;GeV}}{M_\Phi} \right)
\left(
\frac{10^{-3}}{\sqrt{\sum\limits_{\alpha<\beta}|\lambda_{\alpha \beta}|^2 }}
\right)^2 \times (5.0\times 10^{-9} \rm{\;cm}),    
\end{equation}
which means that saturating the bound in Eq.~(\ref{eq:mutoegamma}) leads to $c\tau = 7\times 10^{-11}$~cm for a 100~GeV-mass SCS, which certainly qualifies as prompt. Put another way, the coupling strength that corresponds to $c\tau = 10^{-2}$~cm for $M_\Phi = 100$ GeV is $\lambda = 5\times 10^{-7}$, which is four orders of magnitude smaller than the upper limit from $\mu \rightarrow e \gamma$.  To be clear, the particular pattern of $\lambda$ couplings that we consider here is in no way tailored to evade low-energy constraints.  If only one of the three $\lambda$ couplings is non-zero, flavor-violating signals disappear entirely, and one is left with weaker constraints from flavor-conserving processes. 

As discussed, for example, in Ref.~\cite{Cao:2017ffm}, precision electroweak constraints on an SCS are relatively weak.  Applying the formulas of Ref.~\cite{Grimus:2008nb} to an SCS, one finds that the oblique parameters $S$ and $U$ are never larger in magnitude than $\sim 5 \times 10^{-3}$ even for SCS masses going down to $M_Z/2$, while $T$ does not receive a one-loop correction.    Because corrections to oblique parameters of order a few $\times 10^{-2}$ are consistent with existing data, we conclude that the SCS parameter space is not significantly constrained by precision electroweak data for SCS masses above $M_Z/2$.  

\section{Potential of future BDT-based searches}
\label{sec:BDT}

As the (HL-)LHC accumulates additional data, significant portions of the surviving SCS parameter space will become accessible to an appropriately designed analysis.  Keeping relatively low lepton-$p_T$ and $M_{T_2}$ thresholds will be important in order to be able to probe the low-mass region, $M_\Phi \sim 100$~ GeV. In this section, we investigate the potential sensitivity of an SCS search that uses boosted decision trees (BDTs) and compare it to a cut-based analysis along the lines of that performed in Ref.~\cite{Aad:2019lff}. 

Unlike the recast of Sec.~\ref{sec:recast}, this study requires our own MC simulations of SM background processes. In the fit to data performed in Ref.~\cite{Aad:2019lff}, the four backgrounds $WW$, $WZ$, $ZZ$, and $t{\overline t}$ account for 90\% of the fitted background in the combined SF0J/SF1J/DF0J/DF1J signal regions defined by ATLAS (which have $M_{T_2}>100$~GeV), with the following breakdown: 56\% for $WW$, 6\% for $WZ$, 14\% for $ZZ$, and 14\% for $t{\overline t}$.  We use the same MC pipeline described for the SCS signal in Sec.~\ref{sec:recast} to simulate these four main backgrounds; we relegate further details about our background MC simulations to Appendix.~\ref{sec:MC}.   

Table~\ref{tab:MCcompare} compares signal-region efficiencies for our simulated background events with those based on ATLAS's MC simulations, as inferred from the supplementary materials provided with Ref.~\cite{Aad:2019lff}.  
\begin{table}[htp]
\begin{center}
\begin{tabular}{||c|c|c|c||}
\hline \hline
$WW$ & $WZ$ & $ZZ$ & $t{\overline t}$ \\
\hline 
$(+0.5 \pm 2.2)\%$ & 
$(+25.5 \pm 4.1)\%$ &  
$(+4.8 \pm 3.2)\%$ & 
$(-35.4 \pm 4.4)\%$
\\
\hline \hline 
\end{tabular}
\end{center}
\caption{For the dominant backgrounds, fractional discrepancy of our MC SR efficiencies ($M_{T_2} > 100$ GeV, with SF0J, SF1J, DF0J, DF1J combined) relative to those reported by ATLAS for their MC simulations.  The errors represent the combined statistical uncertainty from our MC and ATLAS's MC.}
\label{tab:MCcompare}
\end{table}%

We find close agreement for $WW$ and $ZZ$, which together account for $\simeq 70\%$ of the background, and larger discrepancies for $t{\overline t}$ and $WZ$.  In the background-only fit to data, Ref.~\cite{Aad:2019lff} includes  rescale factors for these dominant backgrounds as nuisance parameters, with the best-fit values found to be 1.25 for $WW$, 1.18 for $WZ/ZZ$ and 0.82 for $t{\overline t}$. 

\begin{figure*}[t]
\begin{center}
(a) \hspace{0.45\textwidth}
(b) \hspace{0.45\textwidth}

\includegraphics[width=0.45\textwidth]
{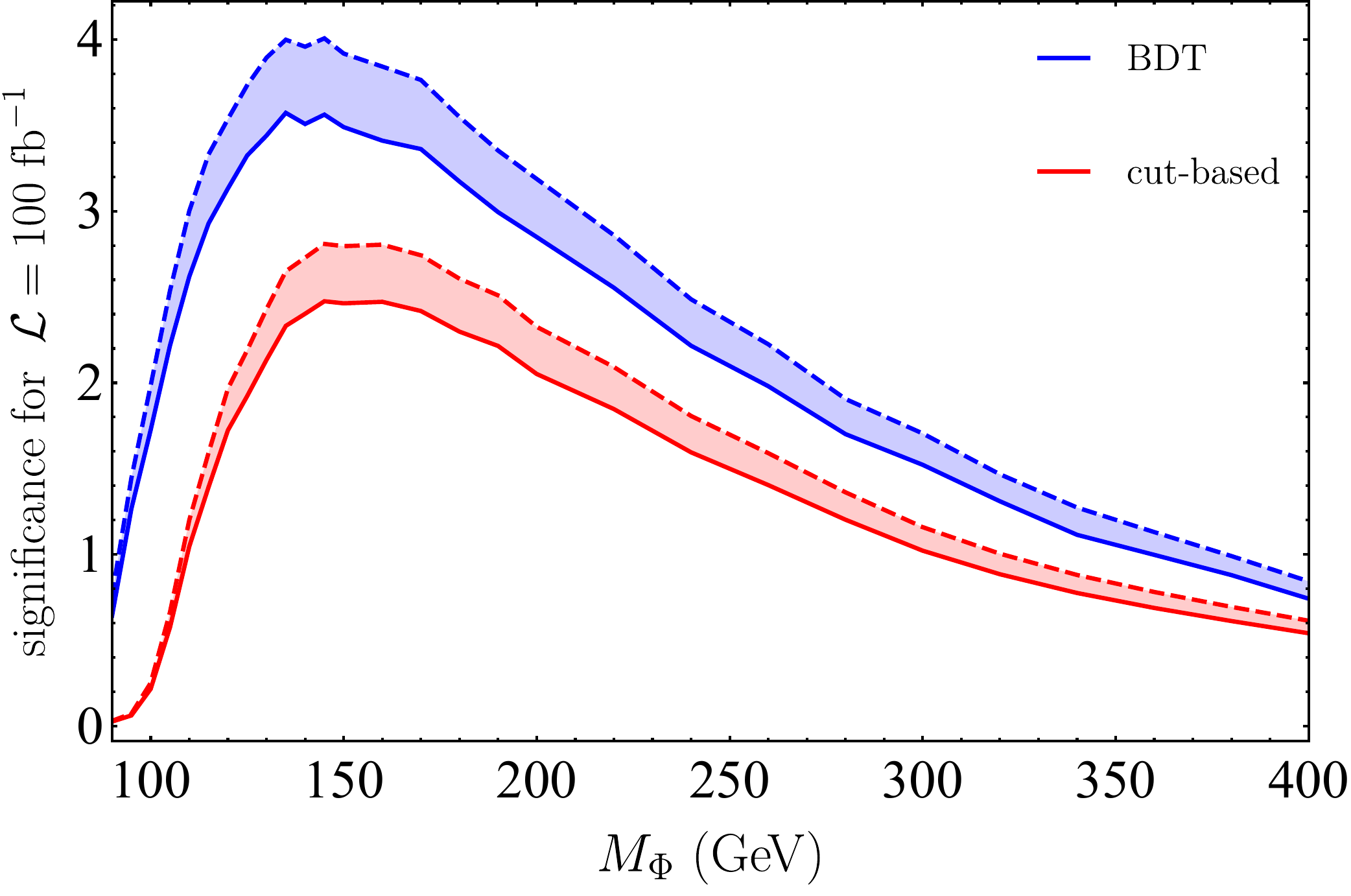}
\;\;\;\;
\includegraphics[width=0.45\textwidth]{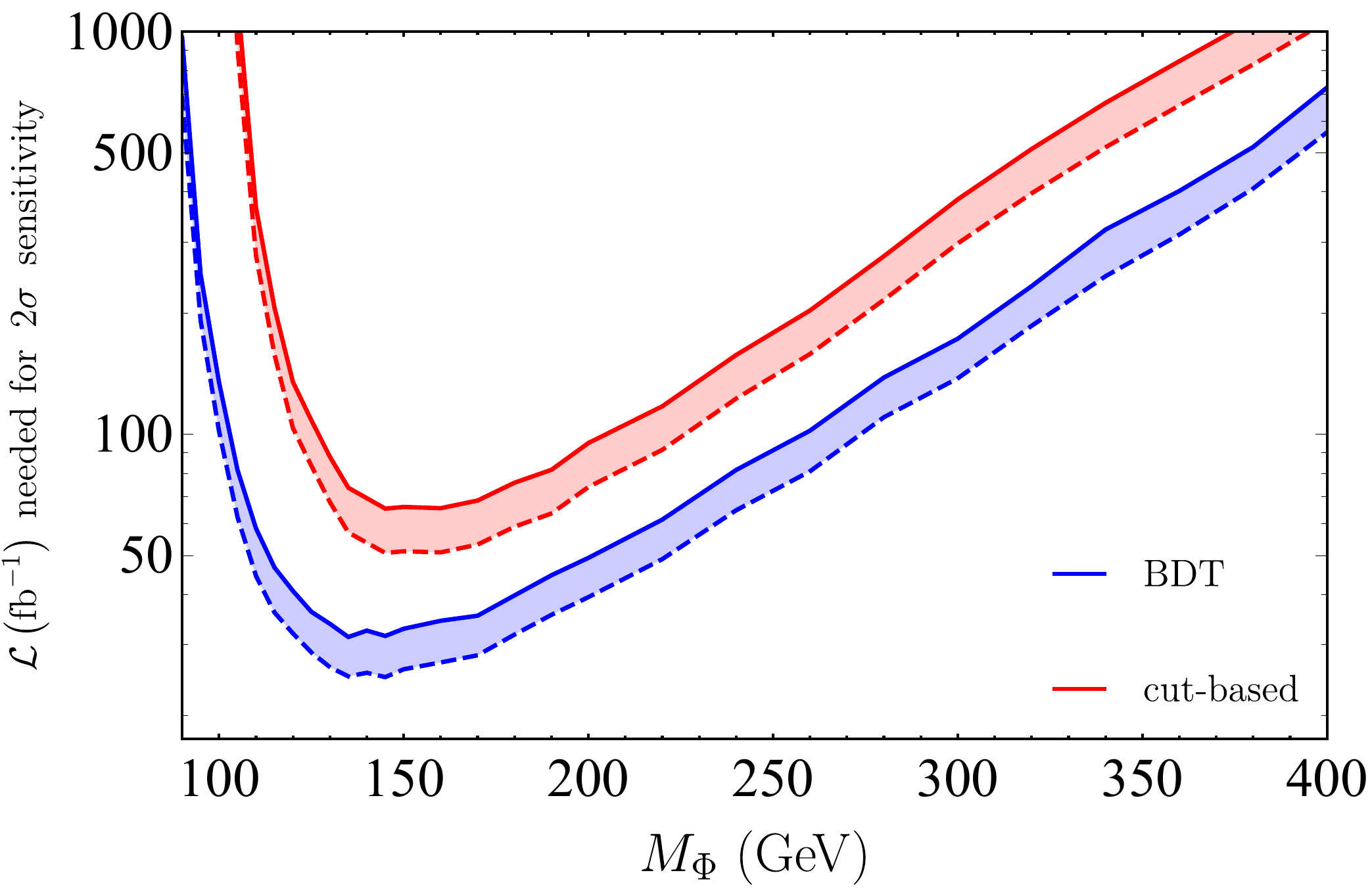}
\\
(c) \hspace{0.45\textwidth}
(d) \hspace{0.45\textwidth}

\includegraphics[width=0.45\textwidth]{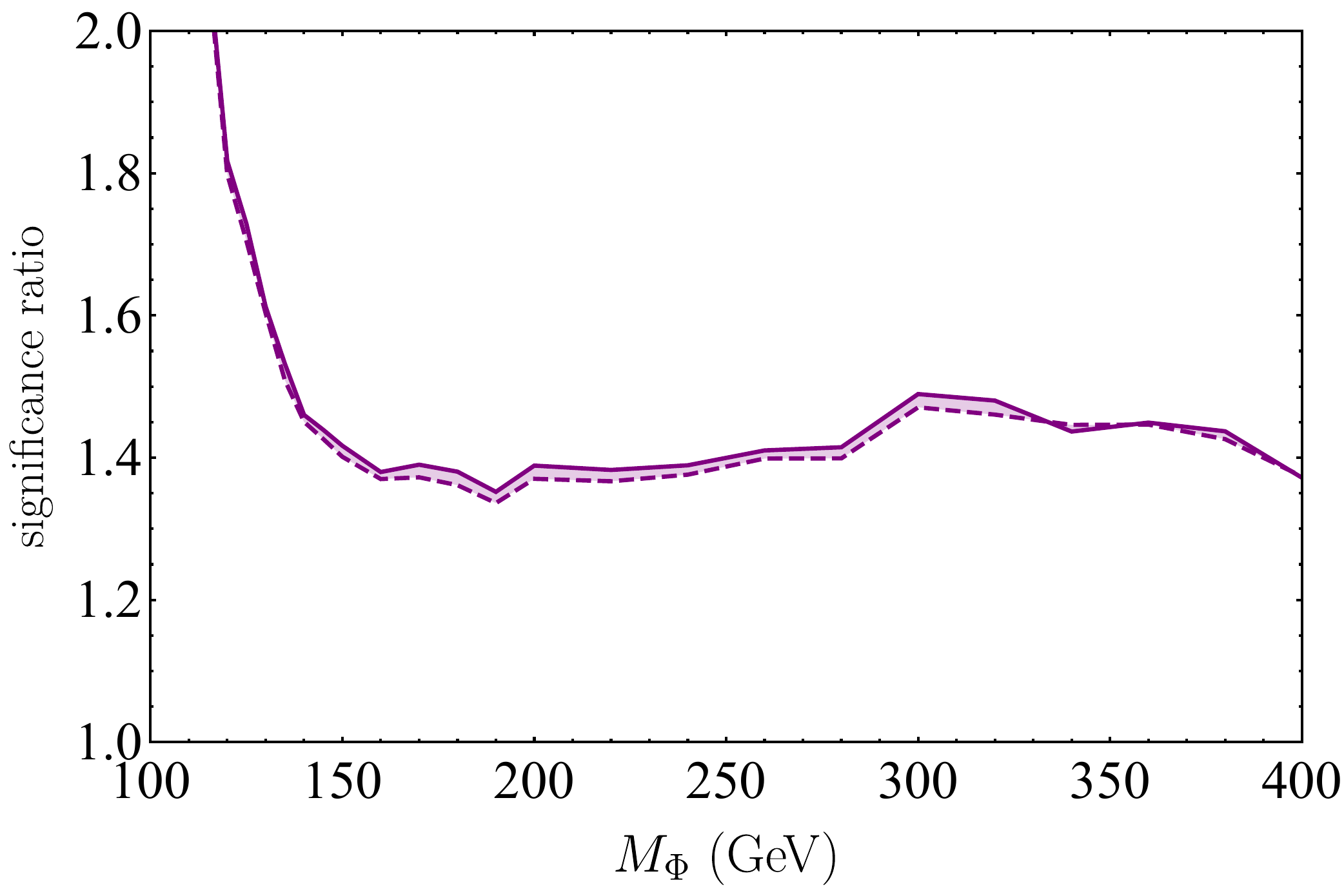}
\;\;\;\;
\includegraphics[width=0.45\textwidth]{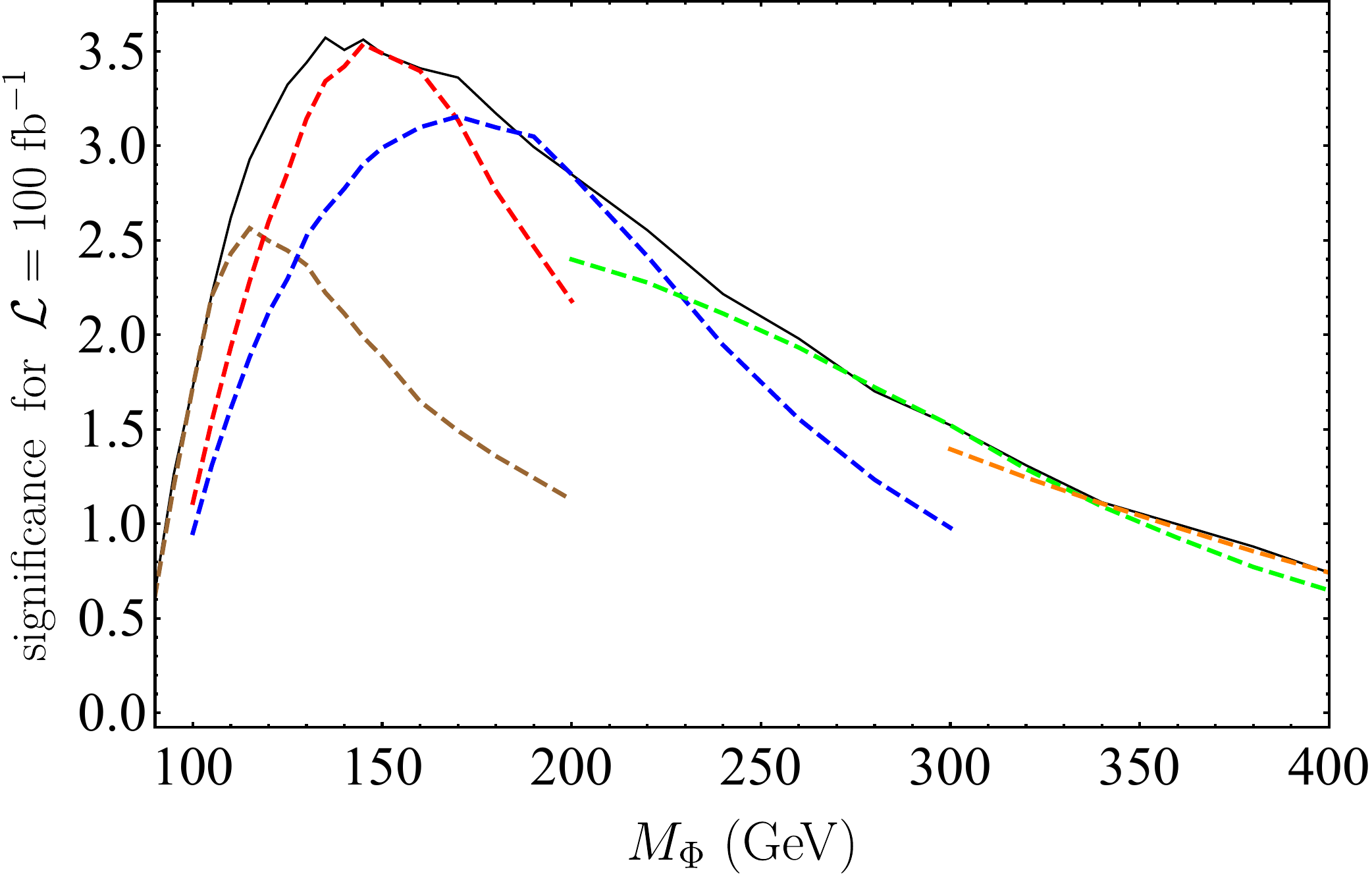}
\caption{
(a):~Expected signal significances, scaled to ${\mathcal L} = 100$~fb$^{-1}$, for the multi-bin  BDT-based and cut-based analyses described in the text.  
(b):~Luminosities required for $2\sigma$ significance.  (c):~The ratio of BDT-based and cut-based significances.
For (a)-(c) the solid (dashed) contours are obtained by rescaling (not rescaling) the weights of our Monte Carlo background samples, based on the results of Ref.~\cite{Aad:2019lff}; (c) shows that the effects of the rescaling largely cancel out when comparing BDT-based and cut-based sensitivities.  (d):~Expected signal significances with the BDT trained on samples with $M_\Phi = 100, 150, 200, 300$, and $400$ GeV (the five dashed curves), and with the BDT trained on the same SCS mass as that of the test sample (solid black; reproducing the solid blue contour of (a)). 
}
\label{fig:multibinBDT}
\end{center}
\end{figure*}

We apply separate reweightings to our four simulated backgrounds, with each reweighting the product of (1) a compensating factor based on the MC discrepancy given in Table~\ref{tab:MCcompare}; (2) the rescaling that ATLAS identifies in their background-only best fit; (3) a common factor $\simeq 1.1$ to compensate for the fact that we neglect single-top and other subdominant backgrounds, estimated by ATLAS to constitute $\simeq 10\%$ of the total background in the combined signal regions. Combining these factors leads to reweightings of 1.38 ($WW$), 1.04 ($WZ$), 1.25 ($ZZ$), and 1.41 ($t{\overline t}$). To keep track of their impact, we present the results of this section both with and without these reweightings.

To develop and test a BDT-based analysis, we use the gradient-boosted implementation in TMVA (BDTG) \cite{hoecker2009tmvatoolkitmultivariate}.  Starting with the events produced by the MC pipeline described in Appendix~\ref{sec:MC}, and again adopting SCS benchmark branching ratios from Sec.\ref{sec:bench},  we first apply the following four preselection cuts:  (1) Exactly two leptons ($e$ or $\mu$).  They must have opposite sign, but in order to obtain results that are less sensitive to the relative sizes of ${\mathcal B}_e$ and ${\mathcal B}_\mu$, we do not distinguish SF/DF. (2) Zero $b$-tagged jets. (3) $M_{ll} > 100$~GeV. (4) $M_{T_2}>85$~GeV.  We train and test a BDT on these preselected events.  We present results for a BDT based on the seven variables $M^{(l_1)}_{T}$ and $M^{(l_2)}_{T}$ (the transverse masses formed using the higher-$p_T$ and lower-$p_T$ lepton, respectively), $n_j$ (the number of jets), $M_{T_2}$, $M_{ll}$, $\Delta \eta_{ll}$, and $\Delta \phi_{ll}$.  This combination is not particularly special; we find that various alternative sets of variables give comparable results.

We bin events according to their BDT score (defined to range from $-1$ to $+1$), setting the bin width to 0.1. With expected distributions for the SCS signal and SM background in place, we use the expected CL$_\text{s}$, which we again calculate with {\tt pyhf}, to quantify potential sensitivity.  Roughly speaking, this amounts to asking how likely it is for the SCS scenario to result in data consistent with SM predictions.  We interpret the expected CL$_\text{s}$ as a double-tailed $p$ value associated with a normal distribution and adopt the corresponding $z$-score as our measure of signal significance\footnote{The significance values that we obtain in this way consistently agree (to within $5\%$) with what we get using $\sqrt{\sum_i S_i^2/(S_i+B_i)}$, where $S_i$ and $B_i$ are the expected signal and background counts in the $i$th bin.}, meaning that 95\%~CL exclusion sensitivity ({\em i.e.} CL$_\text{s} = 0.05$) corresponds to a significance of  $\simeq 2$.  Alongside this BDT-based analysis, we carry out a cut-based analysis along the lines of Ref.~\cite{Aad:2019lff} for the sake of comparison.  This involves the same analysis pipeline that we used for the SCS recast of Ref.~\cite{Aad:2019lff} in Sec.~\ref{sec:recast}, except that we combine SF and DF events to give a total of eighteen bins (nine 0J $M_{T_2}$ bins and nine 1J $M_{T_2}$ bins) instead of thirty-six.  We use expected bin counts for signal and background to calculate signal significance using the same method as for the BDT-based analysis.  

As a function of the SCS mass, expected significances after 100~fb$^{-1}$ are shown in Fig.~\ref{fig:multibinBDT}(a) for both analyses\footnote{For simplicity, we present results in this section only for the $F$-coupled SCS scenario. As discussed in Sec.~\ref{sec:recast}, the $\lambda$-coupled scenario gives slightly stronger signals at higher SCS masses for our SCS benchmark branching ratios.}, with the solid and dashed contours obtained by applying or neglecting, respectively, the background reweightings described earlier. Because we do not attempt to incorporate systematic uncertainties for either analysis, these significances are presumably overly optimistic.  Focusing on the  solid red contour in  Fig.~\ref{fig:multibinBDT}(a), the expected significance is greater than 2$\sigma$ (after 100~fb$^{-1}$) for a range of SCS masses ($\simeq 127-200$~GeV) that is roughly consistent with but somewhat broader than what one might expect based on the recast results of Fig.~\ref{fig:recast_plot} (which are based on a larger integrated luminosity, 139~fb$^{-1}$).  These differences are not surprising, as  Fig.~\ref{fig:recast_plot} gives observed upper limits with background systematics included while Fig.~\ref{fig:multibinBDT} gives expected limits with background systematics neglected. 

Our calculated significances scale as the square root of the integrated luminosity, which allows us to convert Fig.~\ref{fig:multibinBDT}(a) into a plot of the luminosity required to have an expected significance at the $2\sigma$ level, shown as Fig.~\ref{fig:multibinBDT}(b).  In Fig.~\ref{fig:multibinBDT}(c), we see that the ratios of BDT-based and cut-based significances are insensitive to our background sample reweightings.  The BDT-based analysis produces significances that are larger by $\sim 40\%$ for SCS masses above $\sim 150$~ GeV, with more dramatic improvements at lower SCS masses, where it is particularly advantageous to avoid too hard of a cut on $M_{T_2}$. 

With the caveat that an actual experimental search would be subject to systematic uncertainties not included in this comparison, these results suggest that a BDT-based analysis would likely be helpful for probing the most stubborn regions of SCS parameter space. We find that further lowering the $M_{T_2}$ preselection below 85~GeV does lead to further increases in sensitivity at lower SCS masses, although the lower $S/B$ values that result might correspond to larger systematic uncertainties in practice.  All plots in Fig.~\ref{fig:multibinBDT} are based on  simulations of $pp$ collisions at $\sqrt{s} = 13$~TeV.  We expect that the corresponding results for $\sqrt{s} = 13.6$~TeV would be quite similar; for example, the SCS pair-production cross section increases only by $\simeq 9\%$ at $M_\Phi = 100$~GeV and by $\simeq 11\%$ at $M_\Phi = 500$~GeV \cite{susyxsecwg}.

For Figs.~\ref{fig:multibinBDT}(a)-(c), we train a distinct BDT at each SCS mass considered. We can alternatively apply the BDT trained on a particular SCS mass to MC samples with different masses.  In Fig.~\ref{fig:multibinBDT}(d), we take the SCS masses used for training to be $M_\Phi = 100, 150, 200, 300$, and $400$ GeV, and we see that this combined set of BDTs covers the mass range shown reasonably well.

The improved sensitivity of the BDT-based analysis relative to the cut-based one is not tied to the particular binnings we have adopted. In Appendix~\ref{sec:singlebin}, we consider modified analyses in which we simply impose a single optimized cut on the relevant variable (either the BDT variable or $M_{T_2}$), rather than working with binned distributions.  The results of those ``single-bin'' analyses, shown in Fig.~\ref{fig:singlebinBDT}, can be compared with the multi-bin results of Fig.~\ref{fig:multibinBDT}. The significance ratios in Fig.~\ref{fig:singlebinBDT}(c) are again $\simeq 1.4$ or higher, and the gain in sensitivity at lower SCS masses persists even when the $M_{T_2}$ cut is allowed to go below 100 GeV in optimizing the cut-based analysis.

\section{Conclusions}\label{sec:conc}
We have investigated the collider phenomenology of a singly charged scalar that decays promptly to lepton plus invisible. We have seen the challenge of probing the full SCS parameter space:  SCS masses above $\sim 185$ GeV still allowed for the benchmark SCS branching ratios we adopted for our study, along with a lower-mass window around $\sim 100$ GeV.  Appropriately designed dedicated searches leveraging higher integrated luminosity will be able to make significant inroads addressing the surviving parameter space, and they seem well motivated given the relative simplicity of this BSM extension.  Experimental searches should of course aim to probe the SCS parameter space as broadly as possible, including various branching ratio scenarios, and a range of masses for the invisible particle, which we have set to zero here for simplicity.   Moving forward, searches in the $\tau$ and $e/\mu$ channels will continue to be complementary probes of the SCS parameter space.  In the event of a discovery, it will be necessary to search in as many flavor combinations as possible to measure ${\mathcal B}_{e,\mu,\tau}$.  We leave for future work the question of whether $e\tau$ and $\mu \tau$ events might be useful as SCS probes despite the large background from $W$+jets with a fake hadronic $\tau$.   Looking beyond the LHC, future lepton colliders would provide powerful probes of SCS extensions of the SM.   

{\em Note added:}
Near the completion of this work, Ref.~\cite{ATLAS:2025evx} appeared, presenting preliminary results of an ATLAS search that covers new slepton parameter space.  Because the analysis of Ref.~\cite{ATLAS:2025evx} is specially designed  to probe scenarios with small slepton-neutralino mass splitting (including through the requirement of a hard-ISR jet), it does not probe the sensitivity gaps that we have focused on in this paper.

\section*{Acknowledgments}
We are grateful to Jackson Adelman, Sam Bishop, Nathaniel Kirby, Tanvi Krishnan, and Quinn Wilson for work related to this project.  We are grateful to Laura Jeanty and Tracy Slatyer for helpful conversations.   The work of BS is supported by Research Corporation for Science Advancement through Cottrell Scholar Grant \#27632.  The work of DTS  is supported by the U.S. National Science Foundation under Grant PHY-2310770.  This work was performed in part at Aspen Center for Physics, which is supported by National Science Foundation grant PHY-2210452.

\begin{figure*}[t] 
   \centering
   (a) \hspace{0.45\textwidth}
(b) \hspace{0.45\textwidth}

  \includegraphics[width=0.45\textwidth]{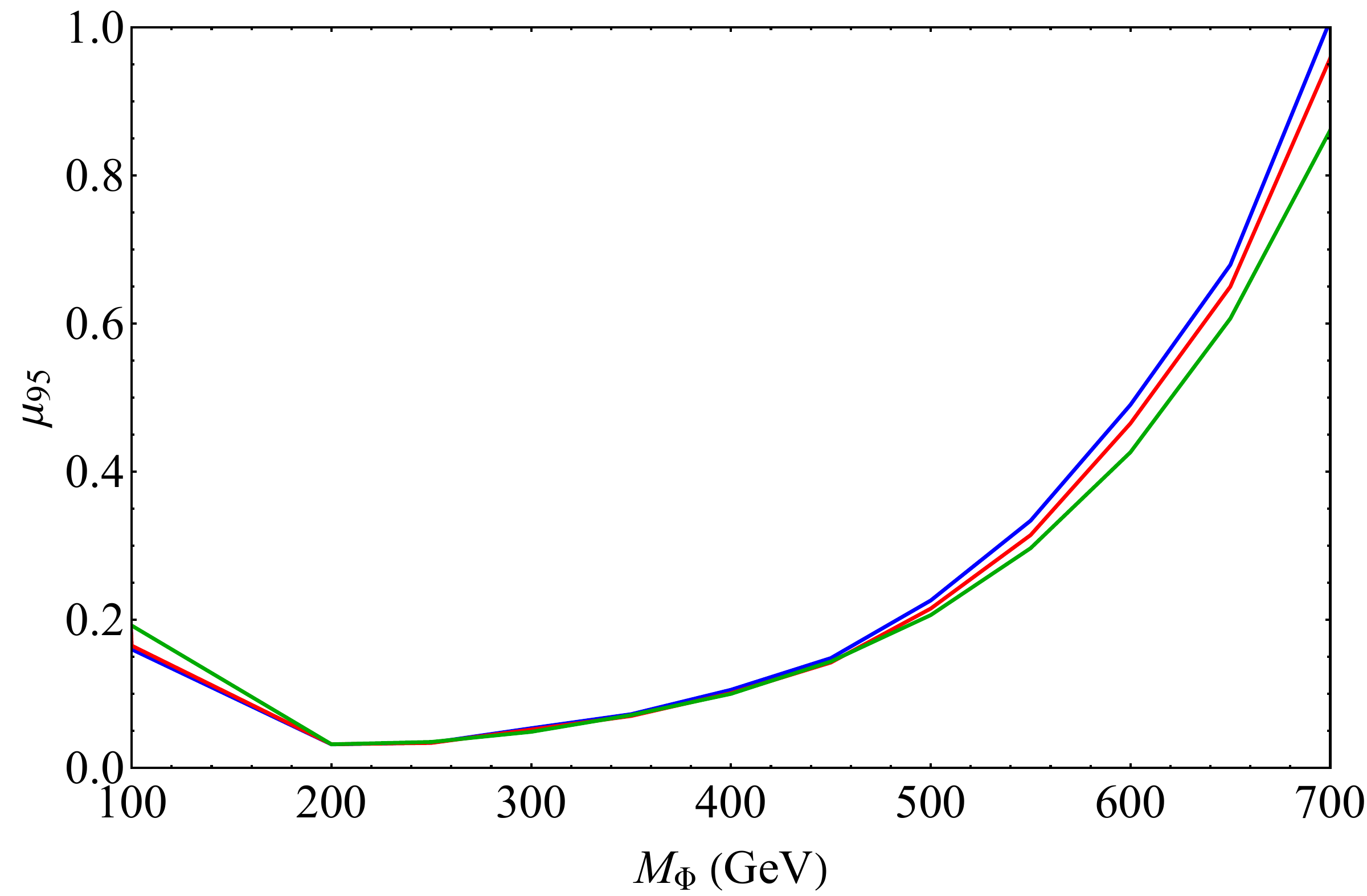} 
       \includegraphics[width=0.45\textwidth]{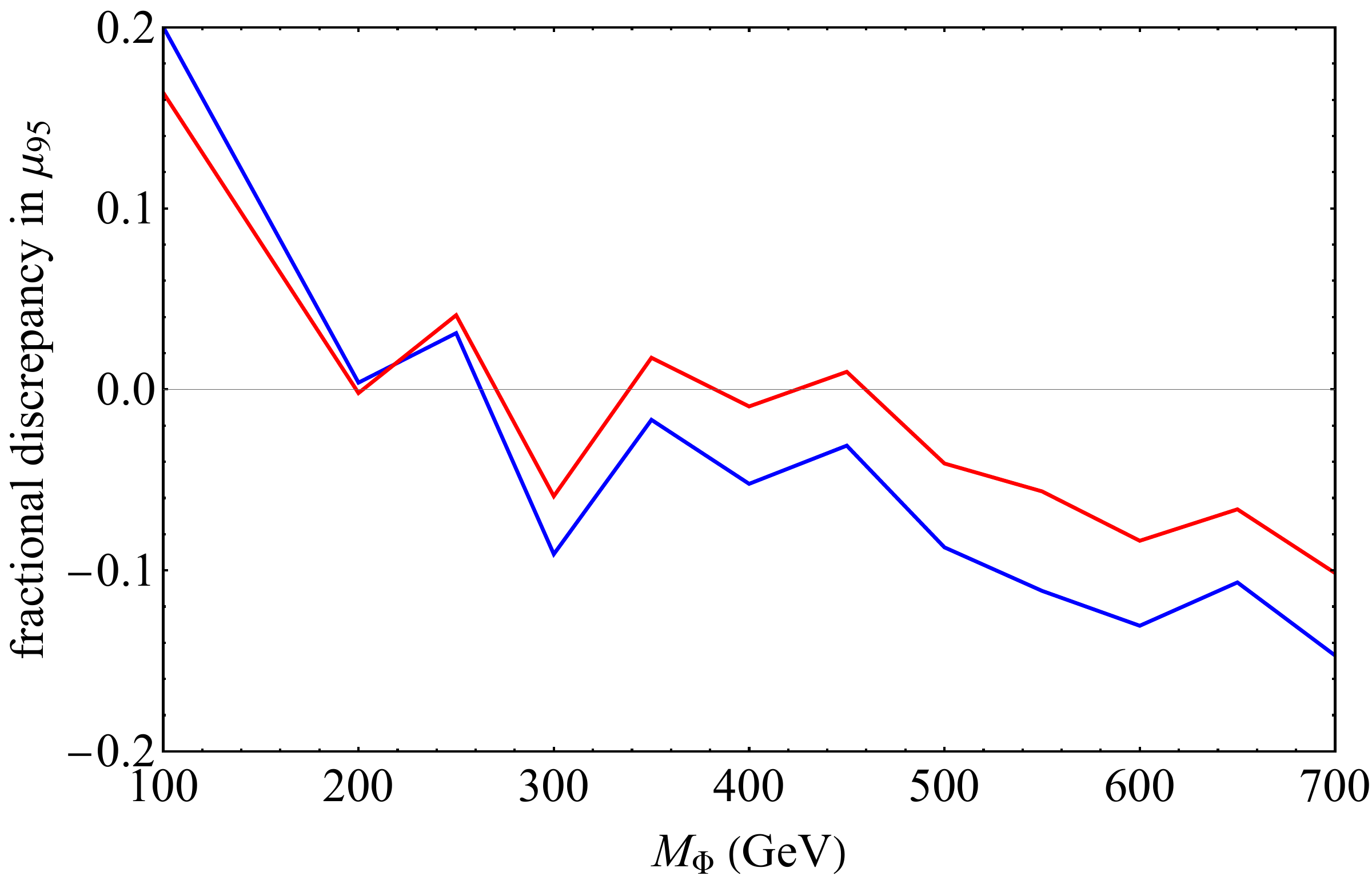} 
   \caption{
(a): Signal-strength upper limits on degenerate sleptons (${\tilde e}_{L,R}$ and ${\tilde \mu}_{L,R}$) as determined using our signal MC (green), using ATLAS's reported MC results including signal uncertainties (blue), and using the ATLAS's reported MC results, neglecting signal uncertainties (red).  (b): the fractional discrepancy of our signal-strength upper limits compared to ATLAS's upper limits including signal uncertainties (blue) and neglecting signal uncertainties (red).  
   }
   \label{fig:validation_slepton}
\end{figure*}
%

\appendix
\section{Monte Carlo event generation}
\label{sec:MC}

In this Appendix we provide details on our MC simulations of the SCS signal and of the $WW$, $WZ$, $ZZ$, and $t{\overline t}$ SM background process.  The SCS reintepretaion of Ref.~\cite{Aad:2019lff} (Sec.~\ref{sec:recast})  uses only our SCS MC samples, while the analyses of Sec.~\ref{sec:BDT} require our background MC samples in addition.     For all simulated processes, we use MadGraph/MadEvent \cite{Alwall:2014hca} for parton-level event generation, Pythia8 \cite{Bierlich:2022pfr} for hadronization and showering, and Delphes \cite{deFavereau:2013fsa} to simulate detector effects.  We adopt MLM jet matching \cite{Mangano:2006rw} with up to two additional jets at the matrix element level.  We set MadEvent run card parameter {\tt xqcut}, which determines the minimum $k_T$ between partons at the matrix-element level, to be 1/3 the mass of the SCS particle for signal simulations, at 30 GeV for $WW$, $WZ$, and $ZZ$ production, and at 60 GeV for $t{\overline t}$ production.  In each case we take the MLM matching scale $Q_{\text cut}$ to be 3/2 times {\tt xqcut}.

As discussed in Sec.~\ref{sec:recast}, our analyses begin with the reconstructed objects returned by a slightly modified version of the code developed by the authors of Ref.~\cite{Araz:2020dlf} to be used within the MadAnalysis framework \cite{Conte:2012fm} in association with a customized Delphes card. For simplicity,  we apply to all events, whether SCS signal or SM backgrounds, the same 85\% trigger efficiency that Ref.~\cite{Araz:2020dlf} applies to slepton events.  The two modifications we introduce are most relevant for the SM background simulations needed for the analyses of Sec.~\ref{sec:BDT}: (1) Consistent with Ref.~\cite{Aad:2019lff}, we count hadronic taus as jets.  (2) Ref.~\cite{Aad:2019lff} reports that for $t{\overline t}$ events,  the algorithm chosen for the ATLAS search has an 85\% $b$-tag efficiency and rejection factors of 2.7, 6.1, and 25 for charm, hadronic $\tau$, and light quarks/gluons, respectively.  We adopt these values throughout, neglecting the dependence on $p_T$, $\eta$, and the underlying process.  

The reconstructed objects returned by our simulation pipeline are electrons with $p_T > 10$~GeV and $|\eta|<2.47$, muons with $p_T > 10$ GeV and $|\eta|<2.7$, and  jets with $p_T > 20$ GeV and $|\eta|<2.4$.  For both the recast of Sec.~\ref{sec:recast} and the analyses of Sec.~\ref{sec:BDT}, we require there to be zero $b$-tagged jets and exactly two leptons (electrons or muons), which must have opposite sign but can either be of the same flavor (SF) or different flavor (DF).

For the recast of  Sec.~\ref{sec:recast} and the ``cut-based'' analysis of Sec.~\ref{sec:BDT}, we further require 
\begin{itemize}
\item either zero jets (0J) or one jet (1J);
\item $p_T > 25$~GeV for both leptons;
\item $M_{ll} > 121.2 (100)$~GeV  for SF(DF) events;
\item $\met > 110$ GeV;
\item $M_{T_2} > 100$ GeV; events are binned in $M_{T_2}$ with the following lower bin-edges in GeV: (100, 105, 110, 120, 140, 160, 180, 220, 260).
\end{itemize}
These are the cuts adopted in Ref.~\cite{Aad:2019lff}, except we do not attempt to simulate ATLAS's object-based MET significance variable \cite{ATLAS:2018uid} and leave out the associated selection.  For the recast in Sec.~\ref{sec:recast}, we have 36 distinct signal-region bins: nine $M_{T_2}$ bins each for SF0J, SF1J, DF0J, and DF1J events.  For the cut-based analysis of Sec.~\ref{sec:BDT}, we combine SF and DF events to more closely mirror our BDT-based analysis, but still distinguish 0J from 1J, leaving eighteen bins.  

For the BDT-based analysis of Sec.~\ref{sec:BDT}, we instead require only $M_{ll}>100$~GeV and $M_{T_2}>85$~GeV, and we bin in the BDT variable, with twenty equal-width bins covering the full range from $-1$ to $+1$. 

When normalizing the weights of our background MC samples we use $\sigma_{t\overline t} = 835$~pb,
$\sigma_{WW} = 119$~pb,
$\sigma_{WZ} = 50$~pb, and
$\sigma_{ZZ} = 17$~pb, 
consistent with NNLO calculations of Refs.~\cite{topxtheo1,topxtheo2,topxtheo3,topxtheo4,topxtheo5,topxtheo6,toppp}
\cite{Gehrmann_2014},
\cite{Grazzini:2016swo}, and
\cite{Cascioli:2014yka},
respectively.  For our SCS signal samples we take the NLO-NLL pair production cross sections for right-handed sleptons from Refs.~\cite{susyxsecwg,Bozzi:2007qr,Fuks:2013vua,Fuks:2013lya,Fiaschi:2018xdm,Beenakker:1999xh}.

In our background MC simulations, we force all $W$ and $Z$ bosons to decay leptonically, with all three lepton flavors included (this includes both $W$'s in $t{\overline t}$ events).  To generate the $ZZ$ sample, one $Z$ is forced to decay to neutrinos and the other is forced to decay to charged leptons, while for the $WZ$ sample the $Z$ is decayed to charged leptons.  By making these choices we effectively require at least two leptons and missing transverse momentum at parton level.  For all four samples we allow one (but not both) of the $W/Z/t$ particles to be produced off-shell, where the kinematic separation between on- and off-shell is set by the value of the relevant MadEvent parameter {\tt bwcutoff}~$15$, meaning the invariant mass of the $W/Z/t$ decay products is restricted to deviate from the pole mass by no more than $\pm15\times \Gamma_{W/Z/t}$.  By off-shell $Z$, we simply mean a $l^+ l^-$ or $\nu {\overline \nu}$ pair with invariant mass far from $M_Z$; in the charged lepton case, both $\gamma$ and $Z$ contributions are included at the amplitude level, with the dilepton invariant mass restricted to $m_{ll}>60$~GeV. For the SCS signal, we assume that the SCS width is small enough to neglect off-shell production and interference between SCS and SM processes.  Our small-width assumption is motivated in part by constraints on lepton flavor violation,   although, as discussed in Sec.~\ref{sec:LFV}, those constraints depend on the flavor structure of the SCS couplings.

\section{Recast cross checks}
\label{sec:recast_check}

As a check on the methods we employ in Section~\ref{sec:recast} for our SCS reinterpretation of Ref.~\cite{Aad:2019lff}, we repeat the exercise for a simplified SUSY model with degenerate sleptons (${\tilde e}_{L,R}$ and ${\tilde \mu}_{L,R}$) and an effectively massless lightest neutralino.  The supplementary {\tt HistFactory} materials provided by ATLAS include expected counts in all signal regions for a variety of slepton masses for this model.  We generate our own MC samples for direct slepton production using the same settings described in Appendix~\ref{sec:MC} for SCS pair production.
We compare the signal-strength upper limits we obtain using our signal MC with the limits that ATLAS's MC gives in Fig.~\ref{fig:validation_slepton}.  The {\tt HistFactory} files provided by ATLAS include systematic and MC-statistical uncertainties for their slepton signals.  To disentangle whether discrepancies are due to these signal uncertainties (which we do not attempt to incorporate for our MC signal), or due to differences in our simulations, we make the comparison with and without the inclusion of the signal uncertainties reported by ATLAS.  Note that the 700~GeV slepton mass lower bound implied by the blue contour in Fig.~\ref{fig:validation_slepton}(a) is consistent with the lower bound reported in Ref.~\cite{Aad:2019lff}.   In the $200-500$~GeV mass range, we see that our signal MC gives agreement with ATLAS results at a better than $10\%$ level. We get a somewhat larger discrepancy, approaching $20\%$, for a slepton mass of 100~GeV.  However, it seems that the ATLAS analysis is nowhere near sensitive to our benchmark SCS model at that mass, as Fig.~\ref{fig:recast_plot} shows.   Unfortunately, ATLAS does not provide information for simplified SUSY models with a near-massless neutralino and slepton masses between 100 and 200~GeV.  

\begin{figure}[t] 
   \centering
   \includegraphics[width=3.5in]{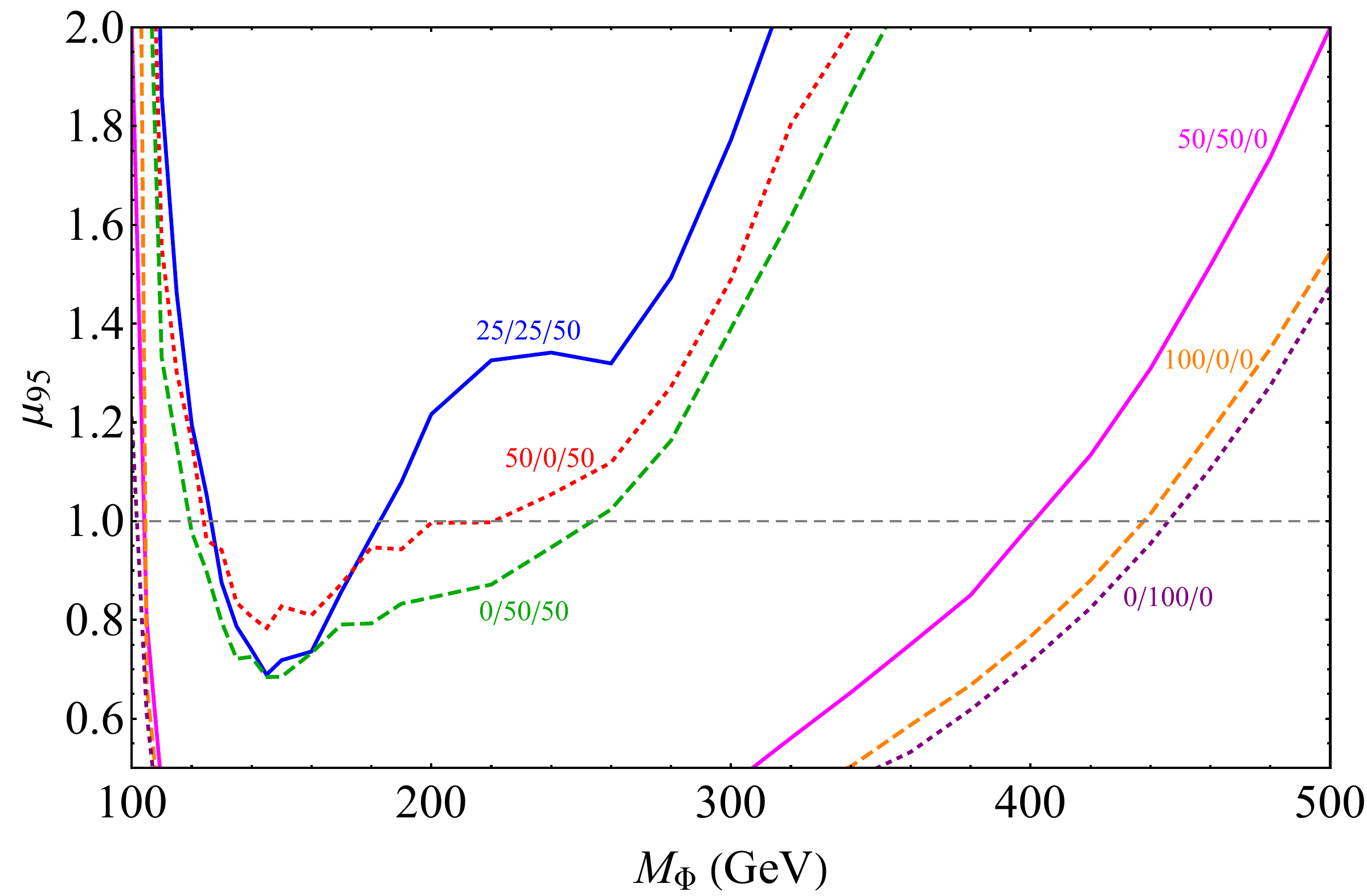} 
   \caption{Reinterpretation of Ref.~\cite{Aad:2019lff} for an SCS that decays through the DM coupling of Eq.~(\ref{eq:DM_coupling}), for various branching ratios ${\mathcal B}_e/{\mathcal B}_\mu/{\mathcal B}_\tau$;  masses with $\mu_{95} < 1$ are ruled out at 95\% CL.  The blue contour coincides with the blue contour of Fig.~\ref{fig:recast_plot}, for our ${\mathcal B}_e = {\mathcal B}_\mu = 25\%$,  ${\mathcal B}_\tau = 50\%$ benchmark.
   }
   \label{fig:recast_various_branchings}
\end{figure}

ATLAS does not include signal efficiencies for right-handed selectrons or right-handed smuons in the {\tt HistFactory} materials that they provide, but they do present separate exclusion contours for scenarios with ${\tilde e}_R$ pair-production and ${\tilde \mu}_R$ pair-production.  Using our signal MC for SCS models with ${\mathcal B}_e = 100\%$ and ${\mathcal B}_\mu = 100\%$, we obtain the ``100/0/0'' and ``0/100/0'' contours in Fig.~\ref{fig:recast_various_branchings}, which give excluded mass ranges $\simeq 105-437$~GeV  and $\simeq 102-446$~GeV, respectively.  These can be compared to the ATLAS exclusions (for a massless neutralino) of $\simeq 117-425$~GeV for ${\tilde e}_R$ and $\simeq 125-450$~GeV for ${\tilde \mu}_R$.  Especially at the lower mass end, the discrepancies are perhaps not surprising given certain simplifying assumptions made in obtaining our estimates; for example (as discussed in Appendix~\ref{sec:MC}), we do not attempt to simulate the MET significance used by ATLAS,  and we apply a flat $85\%$ trigger efficiency to all simulated events, regardless of the SCS mass.  It is also possible that, with the main motivation being to probe the higher mass range, the ATLAS study did not aim for a precise determination of the lower mass boundary, especially given existing constraints from previous slepton searches.  If anything, the discrepancies suggests that the range of allowed lower masses may be underestimated in Fig~\ref{fig:recast_plot}.

\begin{figure*}[t]
\begin{center}
(a) \hspace{0.45\textwidth}
(b) \hspace{0.45\textwidth}

\includegraphics[width=0.45\textwidth]{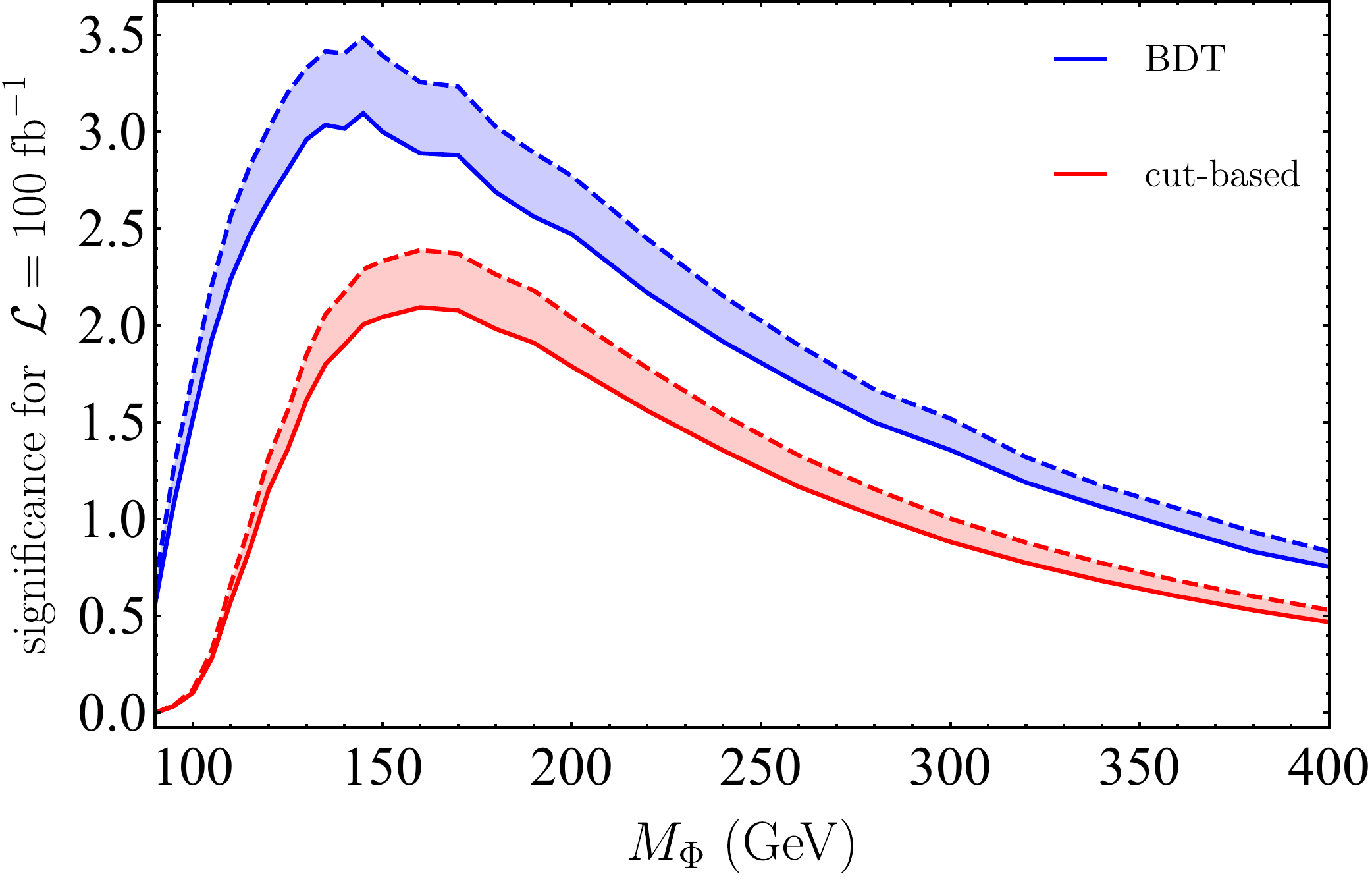}
\;\;\;\;
\includegraphics[width=0.45\textwidth]{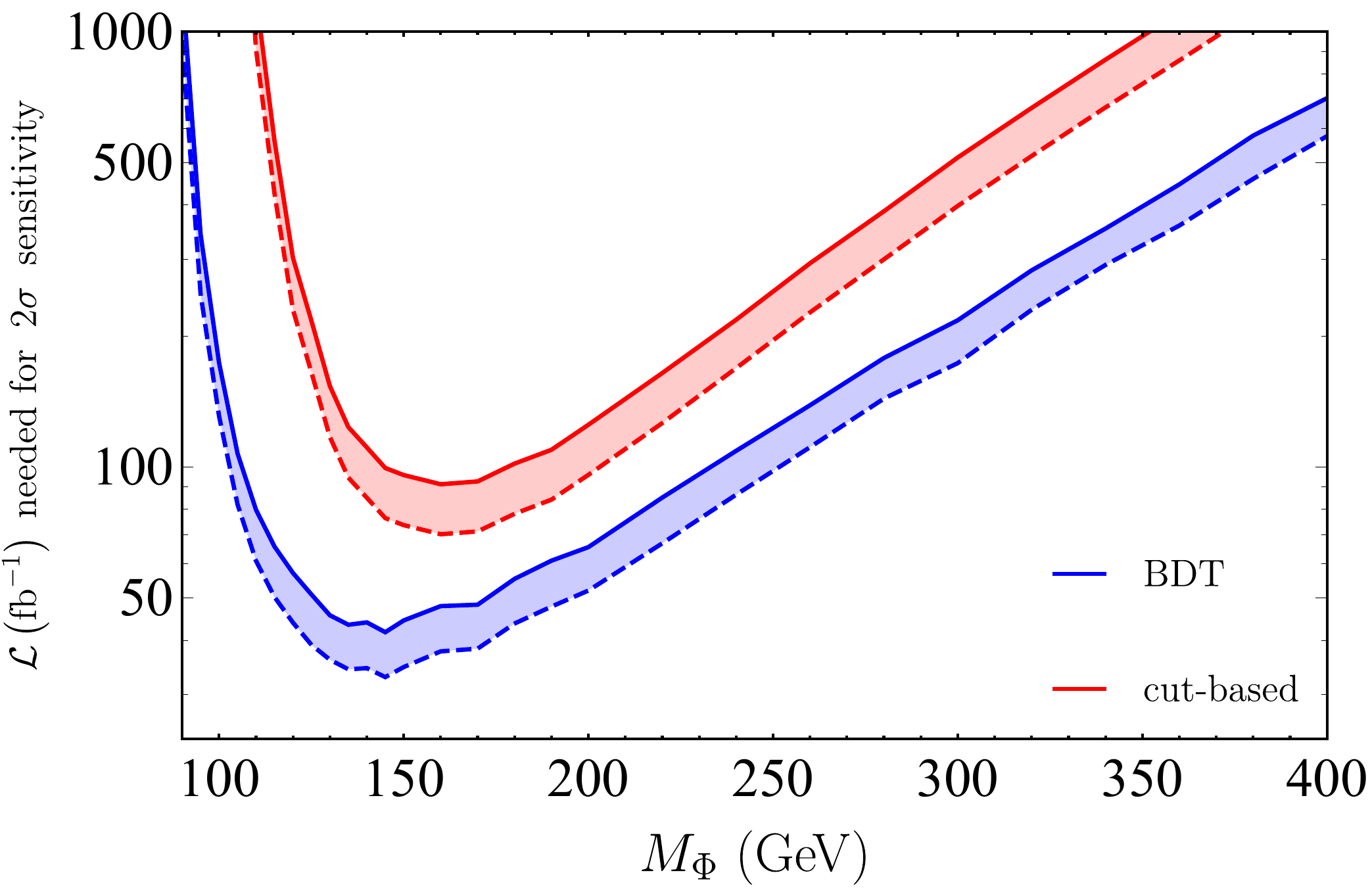}
\\
(c) \hspace{0.45\textwidth}
(d) \hspace{0.45\textwidth}

\includegraphics[width=0.45\textwidth]{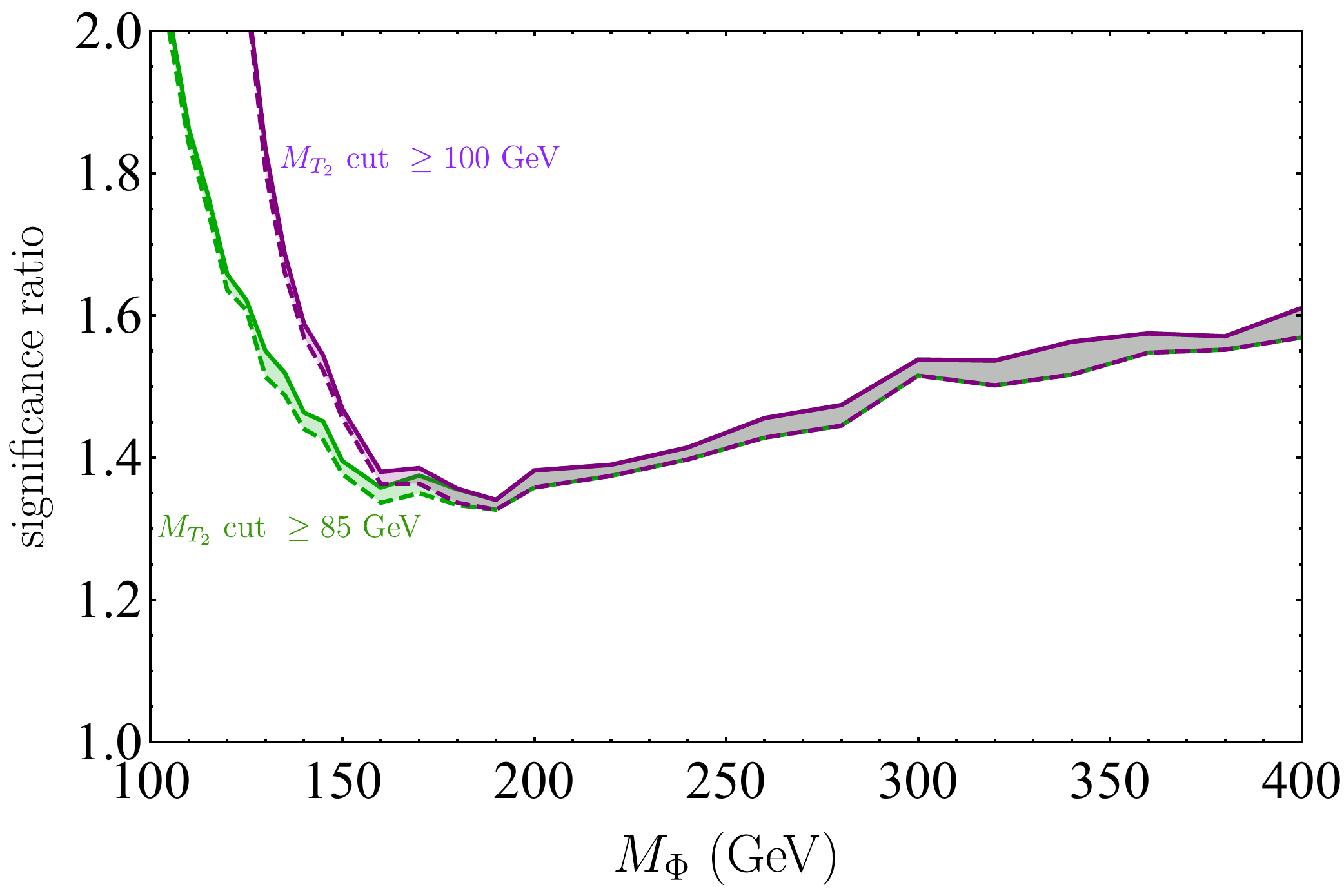}
\;\;\;\;
\includegraphics[width=0.45\textwidth]{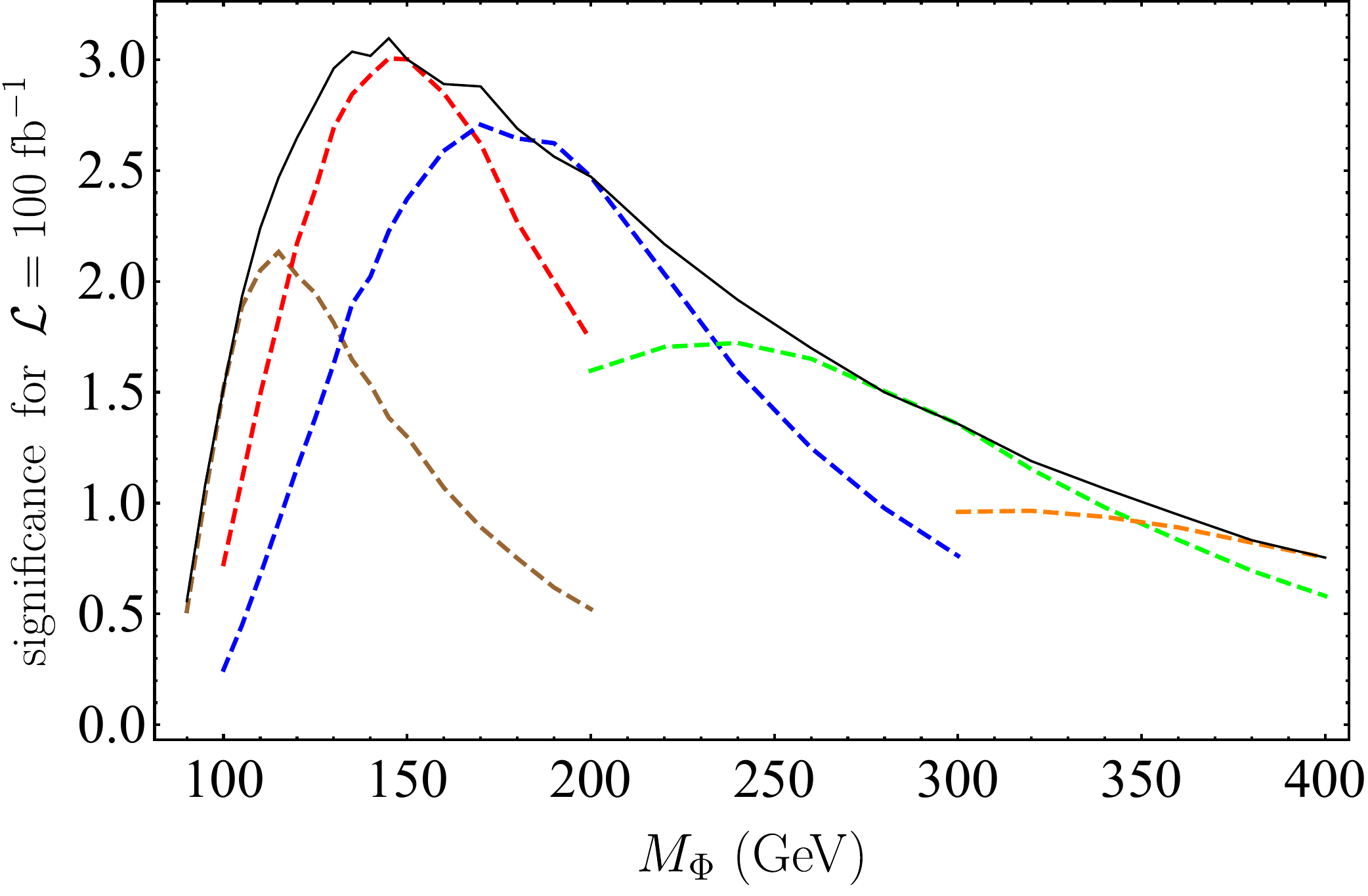}
\caption{Same as Fig.~\ref{fig:multibinBDT}, except with a simple optimized cut on the BDT variable (or on $M_{T_2}$, for the cut-based analysis) instead of a multi-bin treatment.
For the cut-based analysis, we restrict the  $M_{T_2}$ cut to be above 100 GeV; in (c), we show the effect of lowering this cutoff to 85 GeV, which coincides with the preselection cut we adopt for the BDT-based analysis.
}  
\label{fig:singlebinBDT}
\end{center}
\end{figure*}

\section{Recast for other branching ratios}
\label{sec:recast_various_branchings}
 In Fig.~\ref{fig:recast_various_branchings}, we show the results of repeating the recast of Sec.~\ref{sec:recast} with other SCS branching ratios besides our ${\mathcal B}_e/{\mathcal B}_\mu/{\mathcal B}_\tau = 25\%/25\%/50\%$ benchmark.  The auxiliary materials provided by the ATLAS analysis (\cite{Aad:2019lff}) do not break the same-flavor signal regions into $e^+ e^-$ and $\mu^+ \mu^-$ components. For example, it is possible that the bounds on the ${\mathcal B}_\mu = {\mathcal B}_\tau = 50\%$ scenario (represented by the ``0/50/50'' contour in Fig.~\ref{fig:recast_various_branchings}) would be strengthened by restricting to $\mu^+ \mu^-$ signal regions.  Even without leveraging this potential advantage, we see that the excluded mass ranges are broader for the two-flavor scenarios than for our benchmark, extending out beyond $250$~GeV for the ${\mathcal B}_\mu = {\mathcal B}_\tau = 50\%$ scenario.  Nevertheless, a broad parameter space in the few-hundred GeV range and below remains to be explored.

\section{Sensitivity estimates for single-bin analyses}
\label{sec:singlebin}

To demonstrate that our findings are robust to changes in binning, we modify the analysis of Sec.~\ref{sec:BDT} by imposing a single optimized cut on the relevant variable (either the BDT variable or $M_{T_2}$) instead of performing a binned analysis. We show the results in Fig.~\ref{fig:singlebinBDT}.  Comparing Figs.~\ref{fig:multibinBDT} and \ref{fig:singlebinBDT}, we see that the single-bin analyses yield significances that are $\sim 20\%$ lower than for the multi-bin analysis, with very little change to the BDT-based versus cut-based significance ratios. When optimizing the $M_{T_2}$ cut for the cut-based analysis, we impose a 100 GeV lower limit, which coincides with the minimum $M_{T_2}$ value for the signal regions defined by the ATLAS analysis of Ref.~\cite{Aad:2019lff}.  In Fig.~\ref{fig:singlebinBDT}(c), we also show  significance-ratio results when we allow the $M_{T_2}$ cut to go as low as 85~GeV, which coincides with the preselection $M_{T_2}$ cut for the BDT-based analysis.  We see that lowering the $M_{T_2}$ cut does improve the sensitivity of the cut-based analysis at lower SCS masses, but not by enough to erase the sensitivity gains achieved by the BDT-based analysis.       


%

\end{document}